\documentclass[a4paper,10pt,onecolumn]{edpsci} 
\usepackage[]{lineno}
\usepackage[disable]{todonotes}

\newcommand{\textcite}[1]{\citeauthor{#1} \cite{#1}}

\usepackage{float}
\usepackage{mathtools}
\usepackage{array}
\usepackage{amsmath}
\usepackage{booktabs}

\usepackage{graphicx}
\usepackage{epstopdf}
\usepackage[numbers, square]{natbib}
\usepackage{hyperref}
\usepackage{url}
\usepackage{siunitx}

\hypersetup{colorlinks=true,citecolor=blue,urlcolor=blue,linkcolor=blue}

\usepackage[english]{babel}
\usepackage{cmap}
\usepackage[T1]{fontenc}
\usepackage{lmodern}
\usepackage{caption}
\usepackage{subcaption}

\usepackage[ruled]{algorithm2e}
\usepackage{tikz}
\usetikzlibrary{patterns}
\usetikzlibrary{shapes.geometric}
\journalname{Acta Acustica}
\articletype{Research article}

\begin{document}

\title{A Device to Control and Manipulate Occlusion Effects for Own Voice Perception Studies}

\author{
    Rouben Rehman\inst{1}
    \and
    Simon Kersten\inst{1}
    \and
    Aron Schliep\inst{1}
    \and
    Janina Fels\inst{1}
}
\institute{\label{aff1}Institute for Hearing Technology and Acoustics, RWTH Aachen University, Aachen, Germany}

\authorrunning{R. Rehman et al.}

\abstract{
    The occlusion effect (OE) refers to changes of the eardrum sound pressure through ear canal occlusion.
    It consists of two phenomena: an insertion loss (IL) attenuating air-conducted sounds, and an occlusion gain (OG) amplifying bone-conduction.
    Perceptual research on this is hindered by high variability of the OE across individuals, complicating repeatable presentation of precise OE conditions.
    Consequently, a method to control OE conditions reproducibly during perceptual experiments is needed.
    For such investigations, the system must enable separate control of IL and OG.
    We present an approach based on modified commercial earmuffs with integrated microphones.
    The design integrates dedicated impedance measurements and the derivation of digital filters, which are applied to the microphone signals to emulate arbitrary OE curves.
    The system is evaluated objectively through appropriate measurements.
    Results show that the headphones' inherent OG is guaranteed to be below $6$\,dB above $115$\,Hz, falling below $0$\,dB above $155$\,Hz.
    Emulation is shown to work accurately over the entire frequency range of interest.
    Limitations arise mainly due to a slight residual inherent OG for deep voices and the processing delay of the system.
    Future work will focus on the perceptual evaluation of the system and its application in perceptual studies.
}


\maketitle

\section{Introduction}
\label{sec:introduction}
The occlusion effect (OE) is experienced every time one occludes their own ear canals, for example by inserting earplugs or wearing a hearing device. It is characterised by a change in the sound pressure level (SPL) at the eardrums for bodily sounds and the own voice and is defined as
\begin{align}
    \label{eq:initial-oe}
    \text{OE}     = \left|\frac{p^{\text{occl}}_{\text{tm}}}{p^{\text{open}}_{\text{tm}}}\right|, \quad
    L_{\text{OE}} = 20 \log_{10} \left( \left|\frac{p^{\text{occl}}_{\text{tm}}}{p^{\text{open}}_{\text{tm}}}\right| \right)
\end{align}
where $p_{\text{tm}}^{\text{occl}}$ is the sound pressure at the eardrum in the occluded, and $p_{\text{tm}}^{\text{open}}$ is the sound pressure at the eardrum in the open ear case. The OE is well-known to negatively impact satisfaction in users of hearables \cite{dillonHearingAids2012,laugesenOwnVoiceQualities2011}, which is often related to alterations of one's own voice \cite{hengenPerceptionOnesOwn2020}. Furthermore, studies have shown hearables to significantly affect communication behaviour of interlocutors in group conversations \cite{petersenInvestigatingConversationalDynamics2024}. The perceptual principles connecting the dissatisfaction due to the OE with changes in communication behaviour have, however, found only limited attention in research.

\begin{figure}
    \centering
    \includegraphics[width=0.65\linewidth]{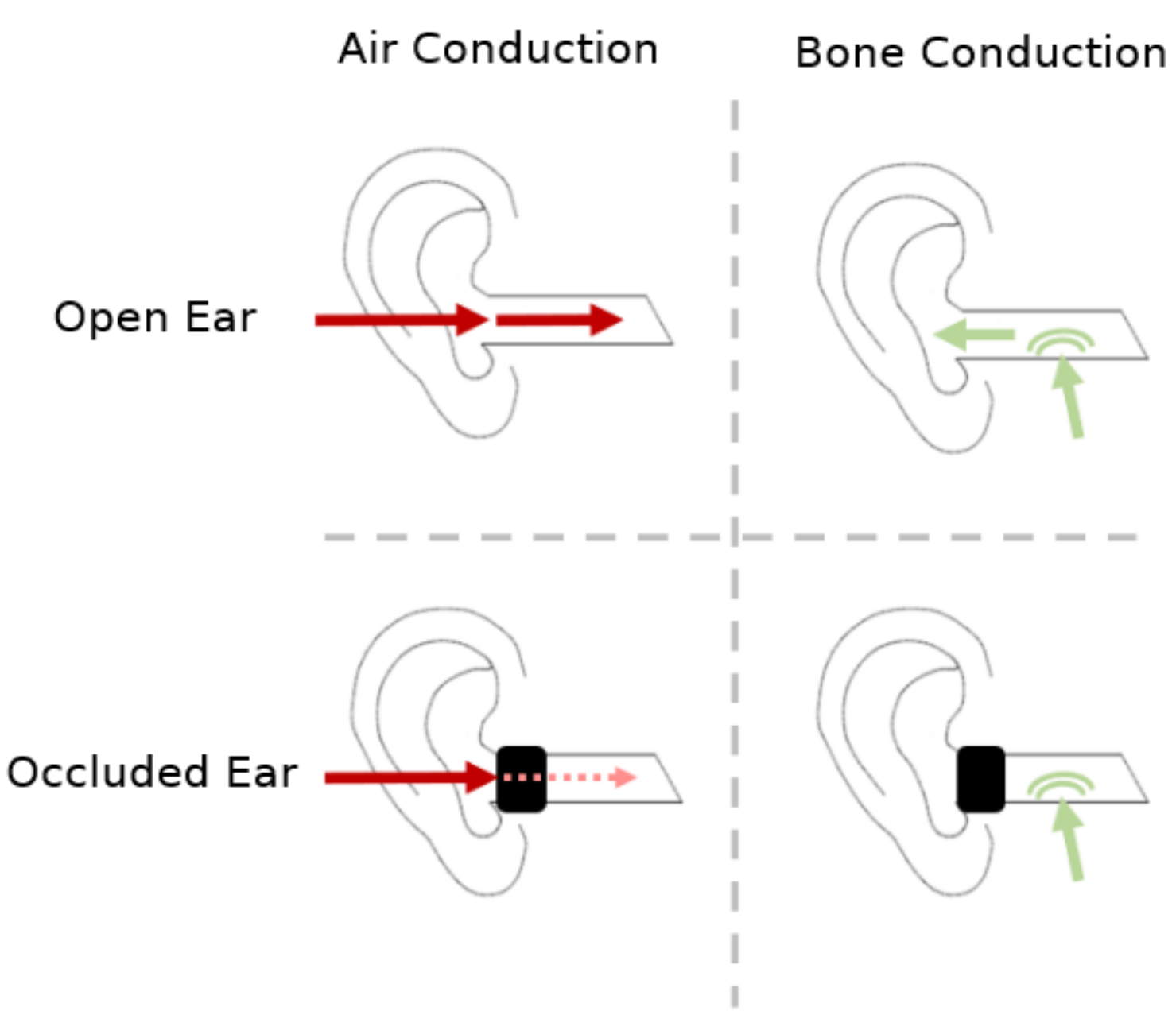}
    \caption{Effects of occlusion on air conduction and bone conduction to the outer ear. Red arrows denote air conduction and green arrows denote bone conduction.}
    \label{fig:ac_bc_vis}
\end{figure}

This is largely due to the complexity of the OE, which makes it difficult to be controlled and systematically investigated in perceptual experiments. Its physical causes are twofold and correspond to the two types of sound transmission applicable to one's own voice. Figure \ref{fig:ac_bc_vis} illustrates both effects. Air conducted (AC) sounds travel from the speaker's mouth through the air into the ear canal and comprise diffraction around the head and reflections from the speaking environment. Bone-conducted (BC) sounds are transmitted via vibrations of the skull bones and surrounding tissues and coupled into the auditory system at the outer, middle, and inner ear \cite{v.bekesyStructureMiddleEar1949,tonndorfNewConceptBone1968}. While occlusion of the ear canal generally affects AC, it is assumed to significantly affect only the BC component to the outer ear. These alterations to AC and BC sounds create amplifications in certain frequency ranges and attenuations in others. The exact ranges and amounts depend on the occlusion device, insertion depth, seal, and the individual ear anatomy, giving the OE its complexity.

Changes to AC sounds are due to the obstruction by the occlusion device, as sounds are attenuated when passing through it into the ear canal. This mainly affects frequencies above 1\,kHz \cite{harrisonMeasurementInsertionLoss2025} and is referred to as insertion loss (IL). For BC sounds to the outer ear, instead amplifications are observed below roughly 1\,kHz. BC sounds are radiated into the ear canal by vibrations of the ear canal walls. At low frequencies and when the ear canal is unobstructed, these dissipate outwards through the ear canal entrance such that the open ear in essence constitutes a high-pass filter characteristic with respect to the tympanic membrane \cite{tonndorfNewConceptBone1968,carilloRemovalOpenEarcanal2021}. Occluding the ear canal prevents this dissipation, thereby causing an increase in SPL compared to the open ear \cite{carilloRemovalOpenEarcanal2021}. In the following, this phenomenon will be referred to as occlusion gain (OG).

In hearing aids, the OG is often mitigated through the use of vented earmolds or open fittings, which allow for some outwards dissipation \cite[Sec. 5.3]{dillonHearingAids2012}.
Deep insertion of the occlusion device has also been shown to reduce low-frequency amplifications. These approaches, however, come with their own drawbacks such as decreased dynamic range, increased risk of feedback \cite[Sec. 5.3.1 \& 5.3.3]{dillonHearingAids2012}, and comfort issues \cite{kiesslingOcclusionEffectEarmolds2005a}. Thus, active occlusion control has been of growing interest in recent years. In this approach, the hearable microphones are used to monitor the sound pressure inside the ear canal. A cancellation signal is then calculated and reproduced to mitigate the OG. While being subject of active research, this approach has been successfully applied in in-ear headphones \cite{denkEvaluationActiveOcclusion2024}, as those most often feature a fully closed design to facilitate active noise cancelling (ANC). In hearing protection on the other hand, meta-earplugs have proven their efficacy in reducing the OG \cite{carilloEnhancingAcousticComfort2025} while also increasing the IL \cite{carilloImprovingLowfrequencyAttenuation2026}. This is achieved through the use of multiple interconnected Helmholtz resonators approximately matching the earplug's load impedance to the open ear radiation impedance.

In perception-focused audiology research, however, the OE is rarely examined. When addressed, it is typically investigated as the consequence of wearing a hearing device and, thus, as a unitary phenomenon without any distinction between OG and IL. A key limitation here is the substantial individual variability making it difficult to establish controlled and reproducible OEs in participants, thereby limiting dose-response investigations. As such, there currently is a disconnect between the practical approaches of OE mitigation and its investigation in perception-focused audiology research. Consequently, the alterations of own voice perception and their implications for voice production and communication behaviour remain largely understudied in both research and standard audiological practice.

To mitigate this, there is a need for a method to precisely control the OE in perceptual studies. This requires individual control of both its OG and IL aspects. In this work, we present our approach using a pair of custom-made closed-back headphones.

Our proposed method is based on the separation of AC and BC sounds by introducing a large earmuff that physically removes AC sounds without inherently creating any significant OG. This allows for a controlled reproduction of the user's voice through active means. To that end, the headphones were built based on a pair of commercially available earmuffs and include microphones on the outside and transducers on the inside to couple the user's speech back into the headphones in real-time. Secondly, we present a method for the derivation of emulation filters given a desired OE target based on a regularised least-squares optimisation. Equalising the setup with these filters allows for the emulation of arbitrary OE targets, where amplifications are bounded by the transducer's dynamic range and attenuations are bounded by the earmuff's overall IL. The system's ability to match the target OE characteristics is evaluated through artificial head measurements.

Similar setups were used in the past to investigate how the own voice is transmitted via bone conduction to the auditory system. \citet{reinfeldtHearingOnesOwn2010} used large earmuffs to quantify the bone conduction transfer to the outer ear for different phonemes by positioning probe tube microphones close to the eardrum and measuring the response during vocalisation. \citet{porschmannInfluencesBoneConduction2000} used similar earmuffs including reproduction hardware to estimate transmission properties of the own voice to the whole auditory system from hearing threshold measurements. These setups, however, were so large and heavy as to require external fixation limiting participants' movements. This would likely interfere with natural communication behaviour, which is the ultimate goal of our research interest, warranting the redesign proposed in this work.

With this work, we aim to build the foundation for future investigations of cause-and-effect relations between the OE and changes in own voice perception, speech production, and communication behaviour. This paper first discusses design requirements for the proposed headphone system, before detailing the methodology used to achieve these targets. Following this, we present a detailed objective evaluation of the system, before concluding with a discussion and outlook on future work.

\section{Methodology}
\label{sec:design}

\subsection{Design Requirements}
\label{subsec:design_requirements}
For the proposed approach, we defined the following set of design requirements:

\begin{itemize}
    \item \emph{Removing Physical Air Conduction:} The headphones have to provide a high level of attenuation for AC sounds, thereby allowing for their controlled reproduction via active means. This is crucial for the emulation of the IL, that starts dominating the OE above ca. 1\,kHz. Including some headroom, we decided on a target attenuation of at least $20$\,dB for frequencies above $500$Hz. This is in accordance with \citet{reinfeldtHearingOnesOwn2010} and comparable to the IL of well-sealing occlusion devices. Between $100$Hz and $500$Hz, at least $15$\,dB of attenuation should be achieved to justify neglecting the leakage component (see Section \ref{subsec:problem_statement}) while keeping system size manageable.
    \item \emph{No Inherent Occlusion Gain:} Matching arbitrary OE targets including a broadband OE of $0$\,dB (open ear case) requires no residual amplification of bone-conducted voice components by the headphones themselves. Thus, they should not introduce any significant OG over the relevant speech-frequency range.
    \item \emph{Occlusion Effect Emulation:} The system must present a plausible, controllable, and repeatable emulation of the OE by matching the OG and IL defined by the OE target independent of the user.
    \item \emph{Real-Time Capability:} The auditory system can detect even minimal delays through comb filtering artefacts \cite{denkDetectionMechanismsProcessing2021a}. Thus, it is crucial to reduce the delay introduced by the signal processing chain to a minimum.  Based on assessments of delay noticeability for external sounds \cite{denkDetectionMechanismsProcessing2021a}, we set a maximum tolerable latency of $10$\,ms. This is also in line with a similar setup from phonology research where participants' speech was altered in real-time \cite{schillerEnhancingAcousticVoice2026} that achieved an "overall system delay of below $16$\,ms".
    \item \emph{Comfort:} To reduce interference with natural communication behaviour as much as possible, the headphones have to be comfortable to wear over longer periods of time without restricting the participants' movements too severely.
\end{itemize}

\subsection{Overall System Design}
\label{subsec:overview}
\begin{figure}[h]
    \centering
    \includegraphics[width=0.65\linewidth]{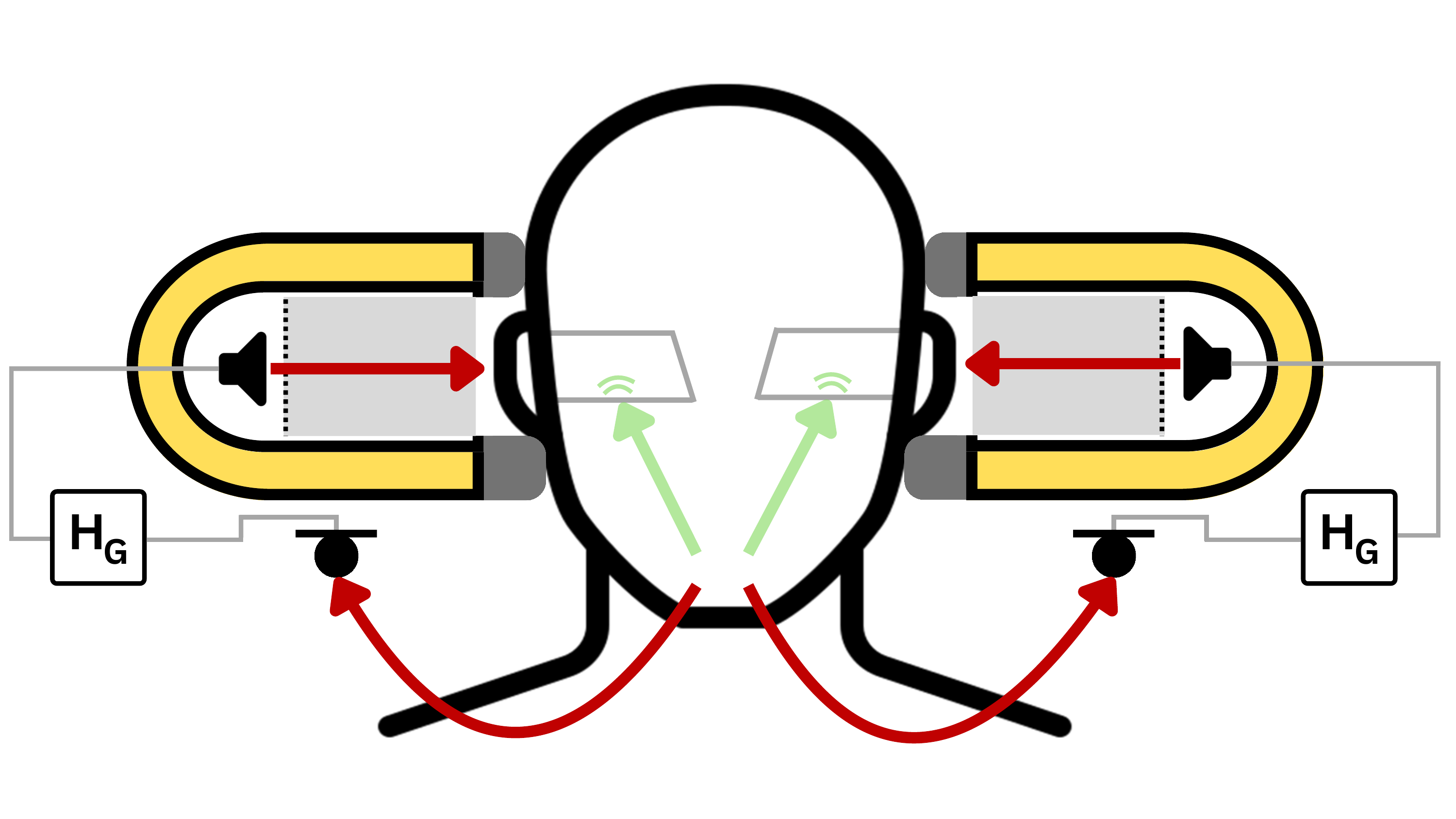}
    \caption{Schematic representation of the described setup. Red arrows denote air conduction and green arrows denote bone conduction.}
    \label{fig:setup}
\end{figure}
\begin{figure}[h]
    \centering
    \includegraphics[width=0.65\linewidth]{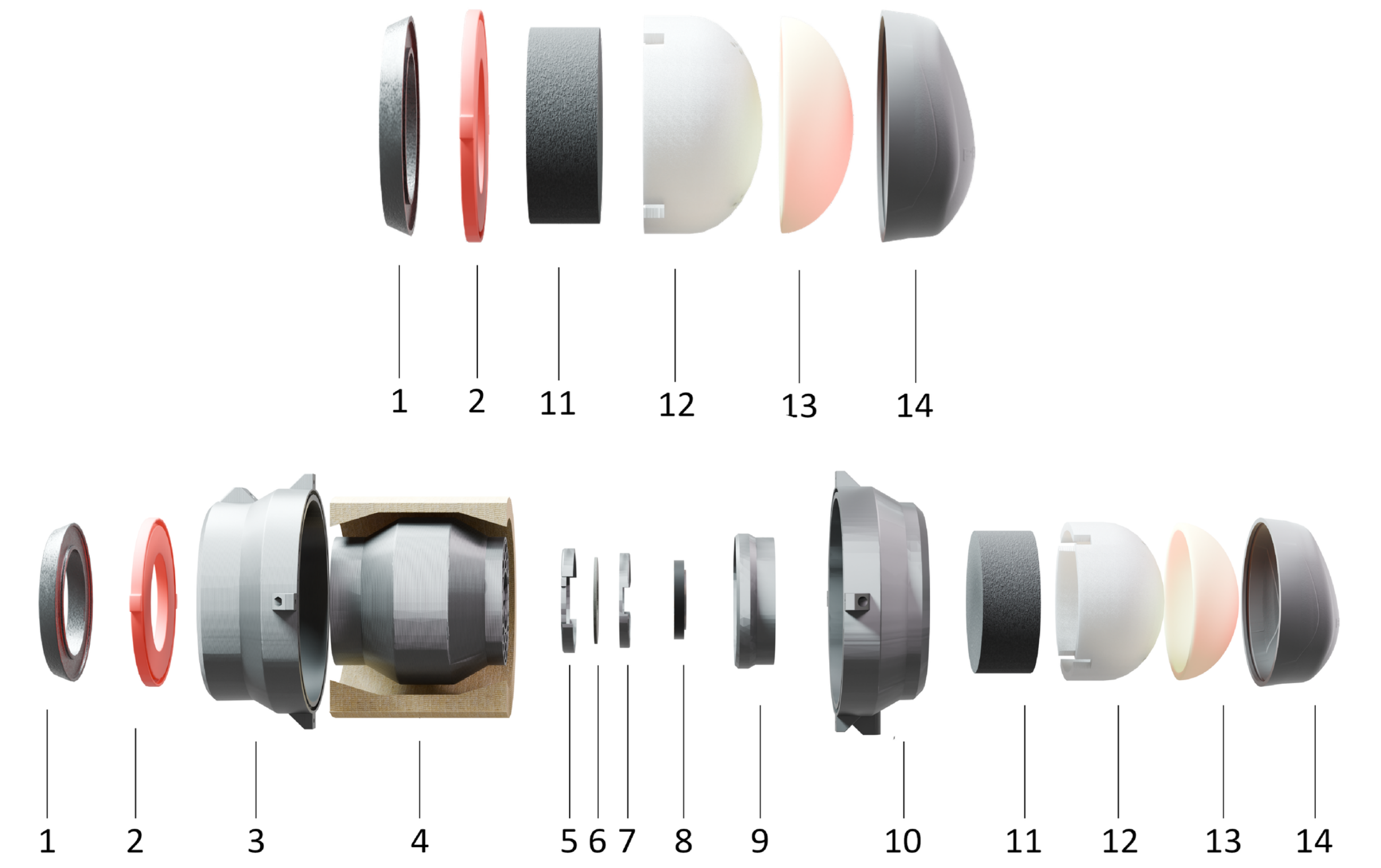}
    \caption{Top: Exploded view of the earmuff capsule in factory condition.\\Bottom: Exploded view of the headphone design.\\Cushioning (1), mounting bracket (2), outer extender shell (3), inner extender shell with mineral wool (4), transducer mounting assembly (5-7), transducer (8), inner extender shell back plate (9), outer extender shell (10), acoustic foam (11), inner plastic cup (12), foam padding (13), rear shell (14).}
    \label{fig:exp_view}
\end{figure}
\begin{figure}[h]
    \centering
    \includegraphics[width=0.65\linewidth]{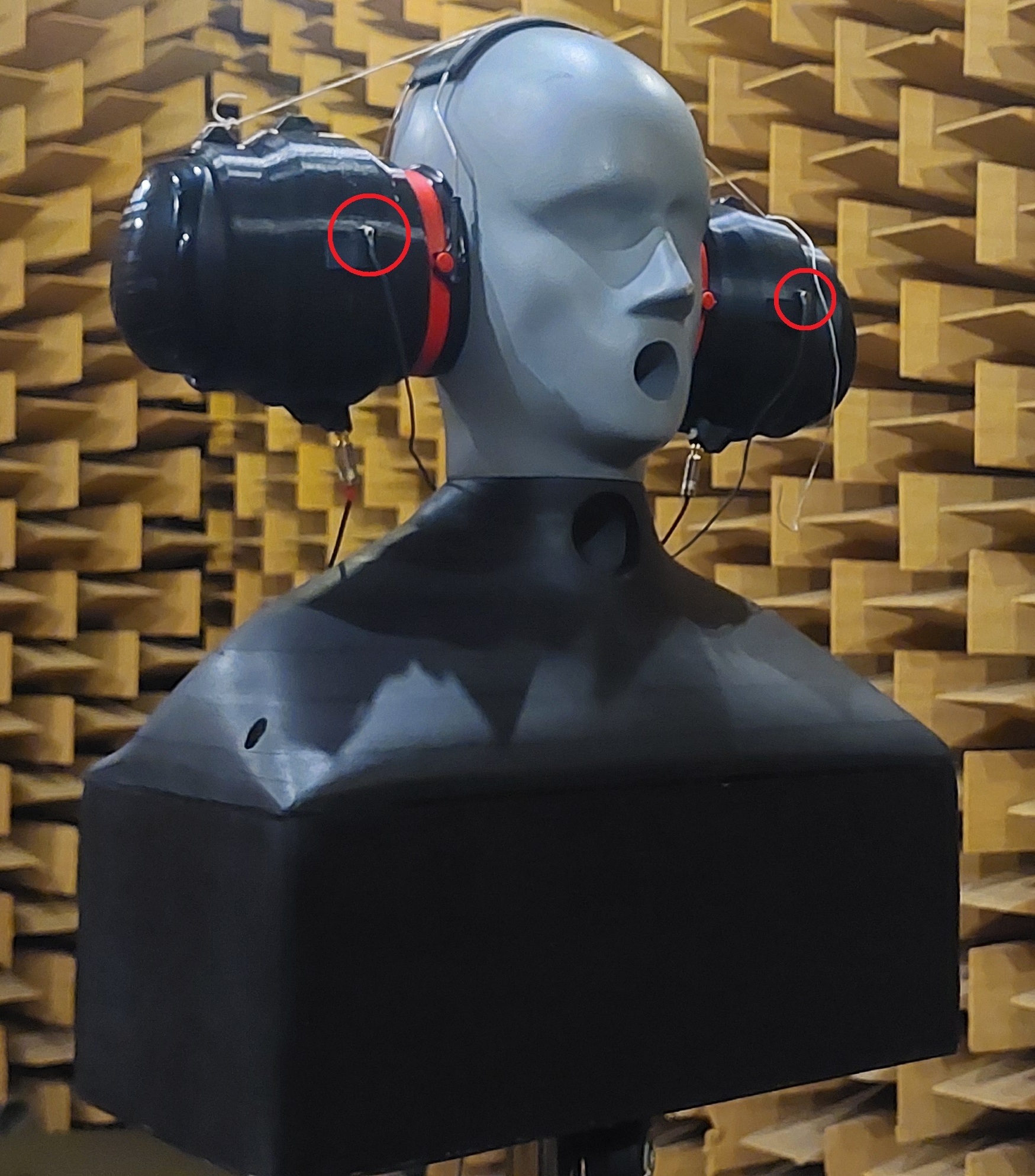}
    \caption{Custom headphones mounted on an artificial head with mouth simulator \cite{kobPhysicalModelingSinging2002}. Microphone positions are marked in red.}
    \label{fig:hp_on_head}
\end{figure}
We based our headphone design on a pair of commercially available earmuffs (3M Peltor Optime III, 3M, Saint Paul, MN, USA). Such earmuffs are specifically designed to provide high levels of sound attenuation while retaining comfort for extended periods of use. Their cups fully enclose the ears and feature a thick, rigid outer shell to minimise sound transmission through the earmuffs \cite{riceDesignFactorsUse1966}.
An important aspect of this is the reduction of leakage to achieve a high system-inherent boundary IL for the emulation and to facilitate neglecting the leakage component in the filter derivation. This is especially critical at the contact area between headphones and head, as the headphones have to accommodate different head sizes and geometries. Figure \ref{fig:setup} shows a schematic representation of the final design, Figure \ref{fig:exp_view} shows an exploded view of the final design, and Figure \ref{fig:hp_on_head} shows the final headphones mounted on an artificial head with mouth simulator \cite{kobPhysicalModelingSinging2002}.

Basing the design on an existing earmuff is beneficial in that regard, as their foam padding is specifically designed to provide a good seal around the ears. Furthermore, the modular design of the earcups allows for easy disassembly and thus enabled us to include our own modifications. We designed custom extender sections to fit between the main mounting bracket (Part $2$ in Figure \ref{fig:exp_view}) and the capsule of the earmuff (Parts $11-14$ in Figure \ref{fig:exp_view}) including a mounting solution for the transducers. The sections and mounting brackets were designed for 3D printing to enable easy replication.

\subsection{Maximising the Inherent Insertion Loss}
\label{subsec:ac_attenuation}
To achieve the target IL, the extender was designed as two independent concentric shells (see Figure \ref{fig:exp_view}). The inner shell attaches to the inner ring of the mounting bracket. It features a mounting armature for the transducer at its other end. The outer shell snaps onto the outer ring on the mounting bracket and extends around the inner shell. On the other end, it accepts the original capsule assembly, causing it to press the foam insert onto the rear of the inner shell, securing it in place. Thereby, no additional rigid connections between the shells were needed, limiting vibrational transmission. Both shells were 3D printed as solid pieces, further hindering sound transmission through the structures. Rather than utilising a standard $100$\,\% solid infill, the extenders were fabricated exclusively from continuous wall loops. This configuration ensures that the extrusion lines remain perpendicular to the incident sound waves, reducing the number of acoustic transmission paths. The space between the shells was filled with mineral wool to further minimise sound transmission.
Due to the narrow-wide-narrow design, the outer shell had to be split into two parts to enable disassembly. To minimise sound transmission through the contact plane, a groove was designed into the cross-section that accepts a rubber gasket providing an effective seal. The inner shell was loosely filled with polyester filling to dampen any standing waves inside the cavity.

\subsection{Minimising the Inherent Occlusion Gain}
\label{subsec:minimize_inherent_bc_change}
In literature, minimisation of the OG was often discussed in terms of the enclosed volume of the earmuffs \cite{porschmannInfluencesBoneConduction2000,reinfeldtHearingOnesOwn2010}. If large enough, this volume will, from the perspective of the ear canal, be essentially indistinguishable from the open ear case. The volumes claimed in literature to fulfil this, however, vary widely. \citet{watsonBoneConductionThresholdMeasurements1943} claimed that, while the effect is only fully minimised for volumes
of $6700$cm³ and above, volumes of $2000$cm³ are sufficient for frequencies down to $100$Hz. \citet{posseltEinflussVerschlussspezifischenKnochenschallHorschwelle1986} and \citet{porschmannInfluencesBoneConduction2000} based their works on this lower boundary, while \citet{reinfeldtHearingOnesOwn2010} settled for a volume of only $245$cm³, stating that this is enough to not cause "an occlusion effect".

Careful consideration, however, shows that the OG is only indirectly tied to the enclosed volume - that is through the load impedance imposed on the outer ear  \cite{carilloRemovalOpenEarcanal2021,rehmanExperimentalSetupEstimation2025}. If the load impedance can be matched to the open ear case, no OG will commence. This is the exact principle behind the OG reduction provided by meta-earplugs \cite{carilloEnhancingAcousticComfort2025}.
Similarly to those previous works, we approached the impedance matching by enclosing a large volume within the headphones.
Formalising the relationship between volume and impedance, however, allows for a more systematic approach to designing such earmuffs:

As the proposed headphones are worn over the ears, they do not physically interfere with the ear canal wall vibrations. Thus, the inherent headphone OG can be expressed in level (see Equation \ref{eq:initial-oe}) as the relation between the total impedances $\underline{Z}_{\text{tot}}^\text{occl}$ and $\underline{Z}_{\text{tot}}^\text{open}$ imposed on the vibrating walls \cite[Eq. 9]{carilloTheoreticalInvestigationLow2020}:
\begin{equation}
    \label{eq:og_level_approx}
    L_\text{OG}=20\text{log}_{10}\left(\left|\frac{\underline{Z}_{\text{tot}}^\text{occl}}{\underline{Z}_{\text{tot}}^\text{open}}\right|\right)
\end{equation}
These impedances are the parallel combinations of the total upstream impedances (towards the ear canal entrance) $\underline{Z}_{\text{up}}^{\text{\{occl, open\}}}$ and the total downstream impedances (towards the tympanic membrane (TM)) $\underline{Z}_{\text{down}}$. 
The downstream impedance in both cases is dominated by the compliances of the residual air volume $C_\text{d}$ and the TM $C_\text{TM}$, hence the superscripts are omitted. For the upstream section in the open ear case, the residual air and ear canal entrance act as masses $L_\text{u}$ and $L_\text{rad}$. Thus, in the low-frequency regime below $1$\,kHz, $\underline{Z}_{\text{tot}}^\text{open}$ is dominated by the upstream section \cite[Eq. 6]{carilloTheoreticalInvestigationLow2020}:
\begin{equation}
    \label{eq:og_approx}
    \underline{Z}_{\text{tot}}^\text{open} \approx j\omega\left(L_{\text{u}} + L_{\text{rad}}\right)
\end{equation}

When wearing the headphones, the residual air mass behind the ear canal entrance and the large internal volume of the headphones form a Helmholtz resonator.
If the impedance of that resonator is significantly smaller than $Z_{\text{down}}$, the approximation from Equation \ref{eq:og_approx} also holds for the headphone case:
\begin{align}
    \label{eq:og_occ_approx}
    \underline{Z}_{\text{tot}}^\text{occl, HP} & \approx j\omega L_{\text{u}} + j\left(\omega L_{\text{load}} - \frac{1}{\omega C_{\text{load}}}\right) \\
                                               & = j\omega L_{\text{u}} + jZ_{\text{load}} \nonumber
\end{align}
By substituting Equations \ref{eq:og_approx} and \ref{eq:og_occ_approx} into Equation \ref{eq:og_level_approx}, we can derive an upper bound for the level of the inherent OG of the headphones that relies solely on the impedances present at the ear canal entrance in the open and occluded case:
\begin{align}
    \label{eq:og_final_approx}
    L_\text{OG} & = 20\text{log}_{10}\left(\left|\frac{\underline{Z}_{\text{tot}}^\text{occl, HP}}{\underline{Z}_{\text{tot}}^\text{open}}\right|\right)           \\
                & = 20\text{log}_{10}\left(\left|\frac{\omega L_{\text{u}} + Z_{\text{load}}}{\omega L_{\text{u}} + \omega L_{\text{rad}}}\right|\right) \nonumber \\
                & < \text{max}\left\{0\text{\,dB}, 20\text{log}_{10}\left(\left|\frac{Z_{\text{load}}}{\omega L_{\text{rad}}}\right|\right) \right\} \nonumber
\end{align}
This representation is beneficial, because it provides an upper bound for the inherent OG of the headphones that is independent of $L_{\text{u}}$ and thus of the individual ear canal geometry. It only depends on the radiation impedance of the ear canal entrance and the load impedance of the headphones, both of which can be measured on a special impedance tube with pinna termination (see Section \ref{eval:inherent-og}).

Before constructing the headphones, we developed a simplified transmission line model for the analytical approximation of the expected load impedance. This was done to inform the design process and achieve the best tradeoff between impedance matching and headphone size. Load impedances for various geometry permutations were calculated and the maximum OG determined for each according to Equation \ref{eq:og_final_approx}. Details on the transmission line model are given in Appendix \ref{app:tl_model}. The final headphone design featured an inner extender with a total length of $11.5$\,cm and a cross-sectional area varying between $30$\,cm² and $59.1$\,cm².
\subsection{Transducer and Microphone Inclusion}
\label{subsec:transducer_inclusion}
Emulating the OE requires the reproduction of the user's filtered speech in real time and thus the inclusion of microphones and transducers. The rear end of the inner extender shell was designed to accept a 40\,mm driver (HPD-40N16PET00-32, Tymphany Acoustic Technology Ltd., Taipei City, Taiwan). This specific driver was chosen, because it was used as a headphone driver in previous scientific work by \citet{mullederUltralightCircumauralOpen2023}, where it demonstrated a satisfactorily flat frequency response when equalised.

Additionally, two MEMS microphones (CMM-2718AT-38164W-TR, Same Sky, Lake Oswego, OR, USA), one for each side, were integrated into the design. To that end, a mounting bracket was designed that slides over the frontal screw posts of the outer extender shell. The shape and tolerances were chosen to ensure precise and repeatable positioning. This allows for easy detaching of the microphones without sacrificing positional accuracy. The microphone positions are marked with red circles in Figure \ref{fig:hp_on_head}.
\subsection{Occlusion Emulation Filter Derivation}
\subsubsection{Mathematical Derivation}
\label{subsec:problem_statement}
Emulating the OE through real-time filtering of the AC voice signal requires appropriate filters designed to replicate the acoustic percept of an occluded ear. For the own voice, the sound pressure at the tympanic membrane is equal to the superposition of its AC and BC components and can hence be expressed as a linear combination of these. In the case of the open ear, \citet{porschmannInfluencesBoneConduction2000} however, already noted that the bone-conducted component to the outer ear has only very little influence on own voice perception.
\begin{equation}
    \label{eq:open-tm-spl}
    p_{\text{tm}}^{\text{open}} = p_{\text{tm, ac}}^{\text{open}} + p_{\text{tm, bc}}^{\text{open}} \approx p_{\text{tm, ac}}^{\text{open}}
\end{equation}
This relation changes in the occluded ear due to the OG. The sound pressure at the tympanic membrane for the occluded ear is thus given by:
\begin{equation}
    p_{\text{tm}}^{\text{occl}} = p_{\text{tm, ac}}^{\text{occl}} + p_{\text{tm, bc}}^{\text{occl}}
\end{equation}
And hence, the OE is approximated as:
\begin{equation}
    OE = \frac{p_{\text{tm}}^{\text{occl}}}{p_{\text{tm}}^{\text{open}}} \approx \frac{p_{\text{tm, ac}}^{\text{occl}}+p_{\text{tm, bc}}^{\text{occl}}}{p_{\text{tm, ac}}^{\text{open}}}
    \label{eq:oe_approx}
\end{equation}
Here and in the following, we omit the absolute value from Equation \ref{eq:initial-oe} as we want to incorporate the magnitude and phase of the OE in the filter design.

To conceptualise the filter derivation, it is necessary to express the tympanic membrane sound pressure using transfer functions relative to a reference point, since the own voice is a subject-internal excitation signal with poor repeatability \cite{blauMethodsExperimentallyCharacterize2025}. A microphone positioned a set distance in front of the mouth is commonly used as such a reference \citep{blauMethodsExperimentallyCharacterize2025}.
Relating $p_{\text{tm}}^{\text{open}}$ and $p_{\text{tm}}^{\text{occl}}$ to this reference sound pressure $p_{\text{ref}}$ individually yields two equations, one for the open and one for the occluded ear, which are also visualised in positions (a) and (b) of Figure \ref{fig:system}.
\begin{align}
    \label{eq:tf_intro_1}
    p_{\text{tm}}^{\text{open}} & = \left(H_{\text{ac}}^{\text{open}}H_{\text{ec}} + H_{\text{bc}}^{\text{open}}\right)p_{\text{ref}} \approx H_{\text{ac}}^{\text{open}}H_{\text{ec}}p_{\text{ref}} \\
    \label{eq:tf_intro_2}
    p_{\text{tm}}^{\text{occl}} & = \left(H_{\text{ac}}^{\text{occl}}H_{\text{ec}} + H_{\text{bc}}^{\text{occl}}\right)p_{\text{ref}}
\end{align}
In Equation \ref{eq:tf_intro_1} (open ear case), we apply the same approximation motivated by \citet{porschmannInfluencesBoneConduction2000} as in Equation \ref{eq:open-tm-spl}. $H_{\text{bc}}$ is the ratio of the BC sound pressure at the tympanic membrane to the reference sound pressure. $H_{\text{ac}}$ is the transfer function from the reference point through the air to the position inside the ear canal, where the medial plane of the occlusion device that is to be emulated would be located in the occluded case. As $H_{\text{bc}}$ and $H_{\text{ac}}$ are affected by the absence or presence of the occlusion device, we use the superscripts \textit{open} and \textit{occl} for differentiation. $H_{\text{ec}}$ is the transfer function from the position of the emulated occlusion device's medial plane to the tympanic membrane. This transfer function is equal in the open and occluded cases, so the superscripts are omitted. Substituting Equation \ref{eq:tf_intro_1} and \ref{eq:tf_intro_2} into \ref{eq:oe_approx} yields the following expression for the OE in terms of the transfer functions:
\begin{equation}
    OE = \frac{H_{\text{ac}}^{\text{occl}}H_{\text{ec}}+H_{\text{bc}}^{\text{occl}}}{H_{\text{ac}}^{\text{open}}H_{\text{ec}}}
    \label{eq:oe_from_tfs}
\end{equation}
From this, we derive an expression for the sound pressure at the tympanic membrane in the occluded ear given a target OE:
\begin{align}
    p_{\text{tm}}^{\text{occl}} & = \left(H_{\text{ac}}^{\text{occl}}H_{\text{ec}} + H_{\text{bc}}^{\text{occl}}\right)p_{\text{ref}}\nonumber \\
                                & = OE \cdot \left(H_{\text{ac}}^{\text{open}}H_{\text{ec}}\right)p_{\text{ref}} \nonumber                     \\
    \label{eq:target}
                                & \eqcolon H_{\text{target}}p_{\text{ref}}
\end{align}
Equation \ref{eq:target} serves as the target function for the filter design. Note, that it only depends on AC transfer functions and the target OE. To artificially emulate the OE, the sound pressure at the tympanic membrane created by the headphones $p_{\text{tm}}^{\text{sys}}$ should resemble $p_{\text{tm}}^{\text{occl}}$. This is achieved precisely when the overall transfer function of the system $H_{\text{sys}}$ matches $H_{\text{target}}$ given a target OE. $H_{\text{sys}}$ is derived through consideration of all sound transmission paths to the tympanic membrane when wearing the headphones, visualised in position (c) of Figure \ref{fig:system}. We obtain:
\begin{figure}[t]
    \centering
    \includegraphics[width=0.65\linewidth]{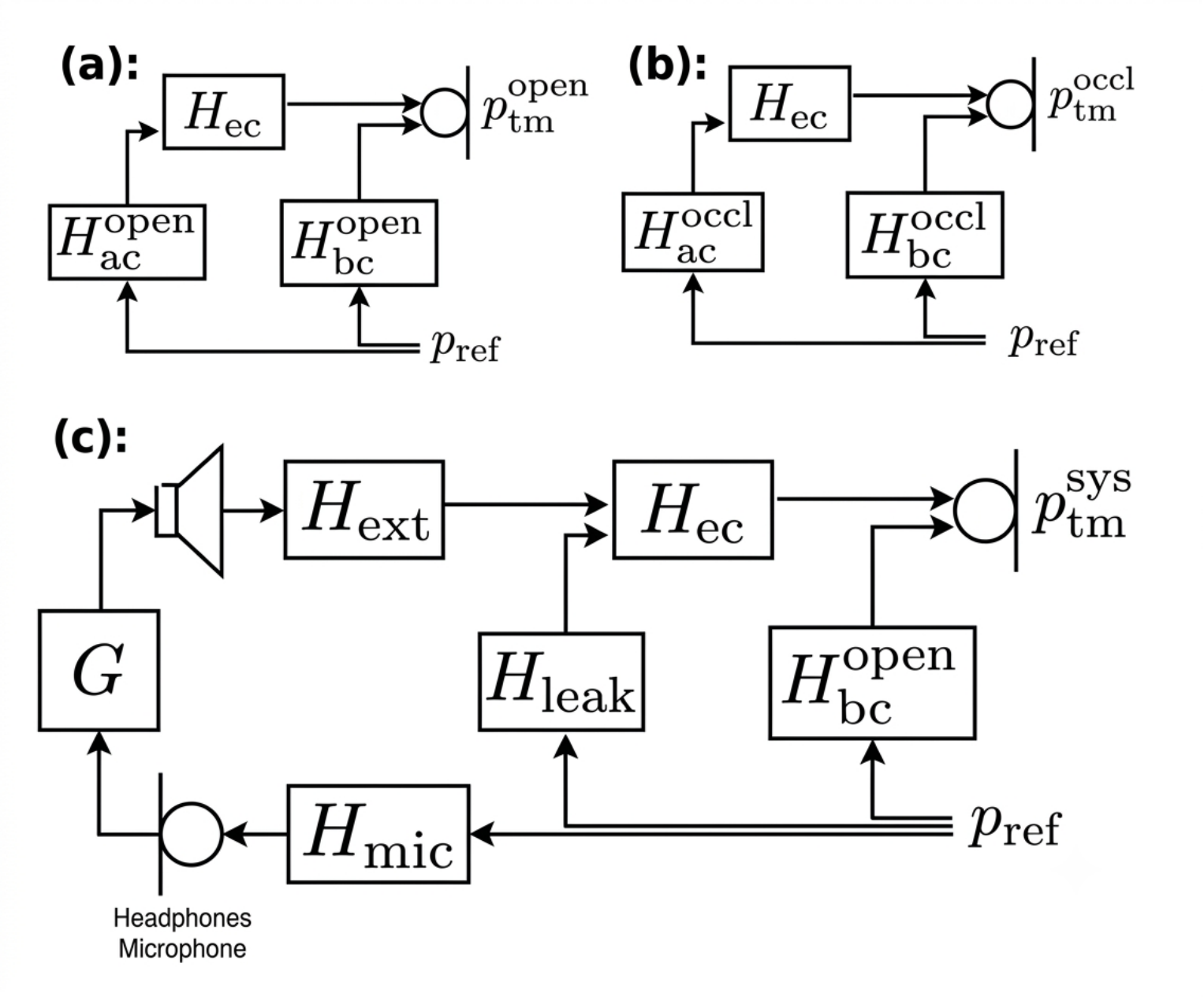}
    \caption{Schematic representation of the open ear (a), occluded ear (b), and headphone system (c) in terms of transfer functions.}
    \label{fig:system}
\end{figure}
\begin{equation}
    H_{\text{sys}}  = \frac{p_{\text{tm}}^{\text{sys}}}{p_{\text{ref}}}
    = H_{\text{bc}}^{\text{open}}+H_{\text{leak}}H_{\text{ec}}+H_{\text{mic}}GH_{\text{ext}}H_{\text{ec}}
\end{equation}
$H_{\text{mic}}$ is the transfer function from the reference point to the microphone, $H_{\text{ext}}$ is the transfer function from the transducer to the position, where the medial plane of the emulated occlusion device would be located in the occluded ear, and $G$ is the filter applied to the microphone signal before it is reproduced by the transducer. $H_{\text{leak}}$ is the transfer function from the reference point to the position of the emulated occlusion device's medial plane in the occluded ear through leakage paths around the headphones. Due to the high passive attenuation of the headphones, this component too is very small compared to the reproduced sound and can thus be neglected. Furthermore, since the headphones are not expected to have any significant inherent OG, we assume the system's BC transfer function to be equal to the open ear case: $H_{\text{bc}}^{\text{open}}$. As discussed above, this is also negligible as it will lie sufficiently below the AC transfer path in level. 
Thus, the equation can be simplified to:
\begin{equation}
    \label{eq:sys_approx}
    H_{\text{sys}}              \approx H_{\text{mic}}GH_{\text{ext}}H_{\text{ec}}
\end{equation}
The goal is to find a filter $G$ that matches the headphone's transfer characteristic to the previously defined target. Equating Equations \ref{eq:target} and \ref{eq:sys_approx} ultimately yields the following solution for $G$:
\begin{equation}
    G = \frac{OE \text{ } H_{\text{ac}}^{\text{open}}}{H_{\text{mic}}H_{\text{ext}}}
    \label{eq:main_G}
\end{equation}
It is important to note that this solution is independent of the bone conduction and ear canal transfer functions. All relevant transfer functions only concern air conduction. Given a target OE, obtaining the appropriate emulation filter $G$ necessitates knowledge of the transfer functions $H_{\text{ac}}^{\text{open}}$, $H_{\text{mic}}$, and $H_{\text{ext}}$ describing the mouth-to-ear, mouth-to-microphone, and transducer-to-ear paths.

Eq. \ref{eq:main_G}, however, is infeasible in practice, as $H_{\text{mic}}$ and $H_{\text{ext}}$ are non-minimum phase, yielding an unstable IIR filter for $G$. To arrive at a practical solution, we employed a regularised least-squares optimisation approach in the time domain.
A similar optimisation approach was previously used for transparency equalisation in hearing aids and hearables by \citet{denkEqualizationFilterDesign2018}, \citet{fabryAcousticEqualizationHeadphones2019}, and \citet{schepkerRobustSingleMultiloudspeaker2022} and has the advantage of directly including the filter length as a constraint.
The formal solution is given in Appendix \ref{app:regul-solution}.

\subsubsection{Obtaining the Required Transfer Functions}
\label{subsubsec:transfer function-meas}
In participants, $H_\text{mic}$ and $H_\text{ac}^{\text{open}}$ can be estimated using
running speech as the input signal similarly to the procedure laid out by \citet{blauMethodsExperimentallyCharacterize2025}. $H_\text{ext}$ in essence constitutes a headphone transfer function (HpTF) and can be directly measured using established methods. As a proof of concept, however, we instead used an artificial head with a mouth simulator and anthropomorphic pinnae which is depicted in Figure \ref{fig:hp_on_head}. This is justified as all necessary transfer functions only concern air conduction and allowed us to measure the transfer functions directly, removing the uncertainties brought about by the estimation based on running speech. Furthermore, this allowed us to use the same setup for the live validation of the system.

The artificial head was designed by \citet{kobPhysicalModelingSinging2002} and features a mouth simulator to accommodate the dependence of $H_{\text{ac}}^{\text{open}}$ and $H_{\text{mic}}$ on the mouth's position and directivity characteristic. The detachable MEMS microphones of the headphone system were used for all measurements. For this purpose, their frequency response was validated to be flat throughout the entire frequency range of interest, spanning from $100$\,Hz to $8$\,kHz, through comparison with a reference microphone with known sensitivity, that was itself equalised to have a flat frequency response (Sennheiser KE3, Sennheiser, Wedemark, Germany).

All measurements were performed using MATLAB (Mathworks, Natick, MA, USA) and the ITA toolbox \cite{ITA-Toolbox_2017} running on a PC (Microsoft Windows 11, Intel Core I7 13700K, 32\,GB RAM). A Yamaha Steinberg UR44C (Yamaha Corporation, Hamamatsu, Japan) was used as the audio interface. All measurements were done inside the same hemi-anechoic chamber. The floor around the measurement setup was covered with additional absorbers to dampen any ground reflections. In addition, all results were appropriately windowed.

For the measurement of $H_{\text{ac}}^{\text{open}}$, the microphones were placed inside the cavum conchae of the anthropomorphic pinnae of the artificial head at the position of the ear canal entrance as they would for an HpTF measurement on a real participant. The measurement was conducted five times and averaged in magnitude and phase individually. The microphones were taken out and repositioned fully for each run to assess the reliability of their positioning.

$H_{\text{mic}}$ was measured by placing the microphones onto the headphones, which were in turn placed onto the artificial head. The measurement was repeated five times, taking off and repositioning the headphones after each run.

Lastly, to measure $H_{\text{ext}}$, the microphones were placed inside the ears of the artificial head again, now with the headphones placed on top. Instead of the mouth simulator, the headphones were now used as the source. The measurement was also repeated five times and averaged in magnitude and phase individually. The headphones were taken off and repositioned for each measurement.
\subsection{Real-Time Capability}
\label{subsec:real-time}
We aimed to keep the overall complexity low to facilitate easy reproduction of the system elsewhere. Consequently, all signal processing was performed on a modern desktop PC (Intel Core i7-14700, 32\,GB RAM) using a Yamaha Steinberg UR44C audio interface (Yamaha Corporation, Hamamatsu, Japan). Live filtering was performed in Reaper (Cockos, San Francisco, CA, USA) using the built-in ReaVerb Plugin. The audio thread priority of Reaper was set to \textit{MMCSS / Time Critical}. Filters were pre-calculated as $256$-tap impulse responses, and the interface buffer size was set to $32$ at a sampling rate of $44.1$\,kHz, as this was the lowest tap length at which the filters could reliably match the target OEs.

\section{Technical Evaluation Procedure}
\subsection{Inherent Insertion Loss}
\label{eval:inherent-il}
Passive attenuation was measured using an ISO 4869-3 \cite{isoAkustikGehorschutzerPart32007} compliant GRAS 45CA test fixture (GRAS Sound \& Vibration, Holte, Denmark). To replicate real-world use, we tested the earmuffs using frontal sound in an ITU-R BS.1116-3 \cite{itu_1116} compliant listening room at RWTH Aachen University. The test fixture was connected to a Nexus conditioning amplifier (GRAS Sound \& Vibration, Holte, Denmark) and positioned on axis 2m in front of the loudspeaker (KH O300D, Klein + Hummel, Ostfildern, Germany) on the same height. Amplifier and speaker were connected to an RME MADIface (RME, Haimhausen, Germany) via an A16 MK-II AD/DA converter (Ferrofish, Linz am Rhein, Germany). The headphones were taken off fully and repositioned after every measurement. Additionally, one measurement was conducted with the empty test fixture for reference. All measurements were performed using the ITA Toolbox \cite{ITA-Toolbox_2017}. The IL was calculated by taking the mean over all headphone measurements, dividing by the reference measurement, and expressing the result in dB. Uncertainty bounds were computed as the mean $\pm$ standard deviation, dividing by the reference measurement, and expressing the result in dB.

\subsection{Inherent Occlusion Gain}
\label{eval:inherent-og}
To evaluate whether the requirement of no significant inherent OG was met, we used an impedance tube with an anthropomorphic pinna termination built by \citet{vorlanderAcousticLoadEar2000}. Since the maximum inherent OG as per Equation \ref{eq:og_final_approx} is independent of $L_\text{u}$, all relevant impedances could be determined through direct measurements. The measurement chain consisted of a Yamaha Steinberg UR44C (Yamaha Corporation, Hamamatsu, Japan) connected to a PC (Intel Core i7-14700, 32\,GB RAM). All measurements and calculations were performed using MATLAB (MathWorks, Natick, MA, USA) and the ITA Toolbox \cite{ITA-Toolbox_2017}. All three microphone positions along the impedance tube were used. A single microphone (Sennheiser KE4, Sennheiser, Wedemark, Germany) was placed in each position with the remaining two being sealed with purpose-built plugs. Two measurements were performed per position and averaged before the impedance was calculated based on the results of all three positions according to ISO 10534-2 \cite{isoAbsorbtionImpedanceTubes2023}. This process was repeated four times for each impedance and the average across all measurements was taken as the final result. The frequency range spanned from $100$\,Hz to $18$\,kHz with the boundaries being dictated by the physical limitations of the transducer.

First, the open ear radiation impedance was measured by leaving the pinna termination open. Then, the load impedance of the headphones was measured by placing them onto the pinna termination. It was assured that the headband of the headphones was under realistic tension during the measurement and the extenders were additionally supported to prevent sagging and ensure a good seal with the artificial pinna. Both sides of the headphones were measured separately. Finally, the obtained open ear radiation impedance and headphone load impedances were compared and used to calculate the maximum OG of the headphones according to Equation \ref{eq:og_final_approx}.

\subsection{OE Emulation}
\begin{figure}[t]
    \centering
    \includegraphics[width=0.65\linewidth]{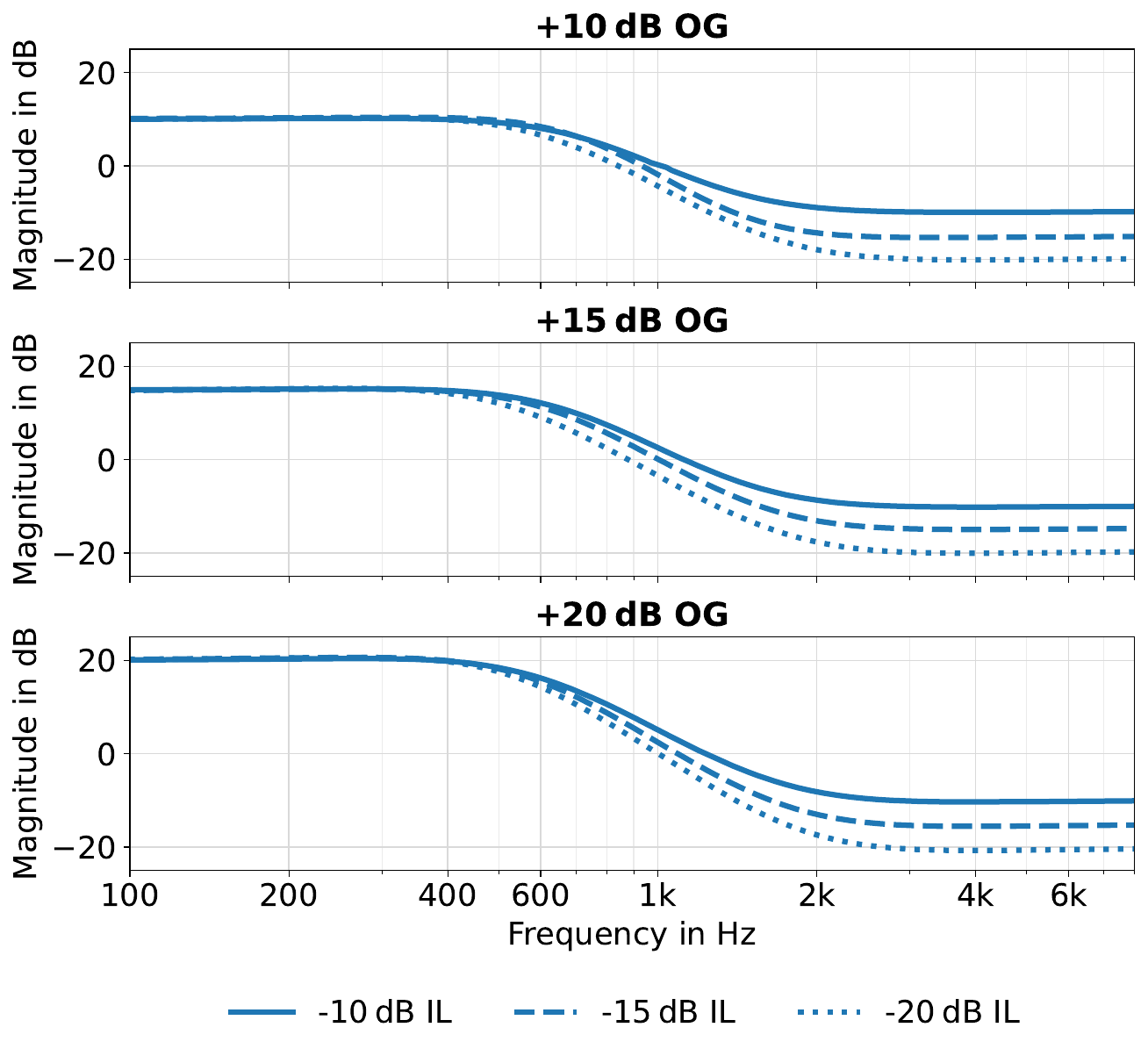}
    \caption{OE target curves with $10$\,dB, $15$\,dB, and $20$\,dB OG and $-10$\,dB, $-15$\,dB, and $-20$\,dB IL used for evaluating the system.}
    \label{fig:oe_targets}
\end{figure}

As the OE is a highly individual phenomenon with large inter-individual variability \cite{blauMethodsExperimentallyCharacterize2025}, we decided to use a set of theoretical OE curves instead of measured ones for the technical evaluation of the system. We designed our OE targets as shelf filters with a centre frequency of 1000\,Hz, placing the upper corner frequency above 400\,Hz. This filter design approximates the theoretical and measured behaviour of vented occlusion devices, which act as a high-cut shelf filter with a cutoff frequency typically between 350\,Hz and 450\,Hz \cite{carilloRemovalOpenEarcanal2021, blauMethodsExperimentallyCharacterize2025}. We included OGs of $10$\,dB, $15$\,dB, and $20$\,dB and ILs of $-10$\,dB, $-15$\,dB, and $-20$\,dB, and built individual OE targets for each combination to prove the system can independently control both. Additionally, we included a flat $0$\,dB target, yielding a total of $10$ OE targets. All targets were designed to have minimum phase.
Their magnitude spectra are depicted in Figure \ref{fig:oe_targets}.
With these OE targets, the corresponding target transfer functions for the headphones were derived according to Equation \ref{eq:target}. The necessary emulation filters were then calculated using the regularised least-squares optimisation approach (see Appendix \ref{app:regul-solution}) for a tap length of $256$.

To investigate the system's ability to match the target transfer functions, the artificial head with mouth simulator described in Section \ref{subsubsec:transfer function-meas} was used. Live filtering was facilitated by loading the emulation filters into the Reaper (Cockos, San Francisco, CA, USA) setup described in Section \ref{subsec:real-time}. One MEMS microphone was placed inside the cavum conchae of the artificial head, while the other was mounted on the headphones, which were then placed on the head.
The impulse response measurements were conducted in MATLAB (Mathworks, Natick, MA, USA) with the ITA Toolbox \cite{ITA-Toolbox_2017} while the real-time filtering was active. The live filtering ran simultaneously on a parallel machine. Specifically, the headphone transducer and microphone were routed through an additional interface (Yamaha Steinberg UR22C, Yamaha Corporation, Hamamatsu, Japan) connected to a separate PC (Microsoft Windows 11, AMD Ryzen 7 5700U, $16$\,GB RAM).

To ensure playback of the filtered signal at the correct level, the $0$\,dB target filter was activated and the system's output amplification adjusted until it matched the level of $H_{\text{ac}}^{\text{open}}$ (see Section \ref{subsubsec:transfer function-meas}). Afterwards, the emulation result for each OE target was measured three times, referenced to a microphone positioned $100$\,mm in front of the artificial mouth, and averaged in magnitude and phase individually. The resulting measurements of the headphones' transfer characteristic under the emulation filters were then compared to the ideal targets derived from Equation \ref{eq:target}.

\subsection{Real-Time Capability}
To evaluate real-time capability, round-trip latency was measured by placing the headphones on a GRAS 45CA headphone test fixture, which was connected to the Yamaha Steinberg UR44C audio interface (Yamaha Corporation, Hamamatsu, Japan) via a Nexus conditioning amplifier (GRAS Sound \& Vibration, Holte, Denmark). A loudspeaker (Genelec 6010A, Genelec, Iisalmi, Finland) was positioned on axis $2$m in front of the test fixture and the measurement performed in  MATLAB (Mathworks, Natick, MA, USA) using the ITA Toolbox \cite{ITA-Toolbox_2017} while the real-time filtering was active on an auxiliary machine. The same measurement was then repeated with the headphones and real-time processing removed. The round-trip latency was determined as the time difference between the peaks of the two measurements and was found to consistently be around $6.5$\,ms.

\section{Results and Discussion}
\label{sec:results}
\subsection{Inherent Insertion Loss}
\label{subsec:res_IL}
\begin{figure}[t]
    \centering
    \includegraphics[width=0.65\linewidth]{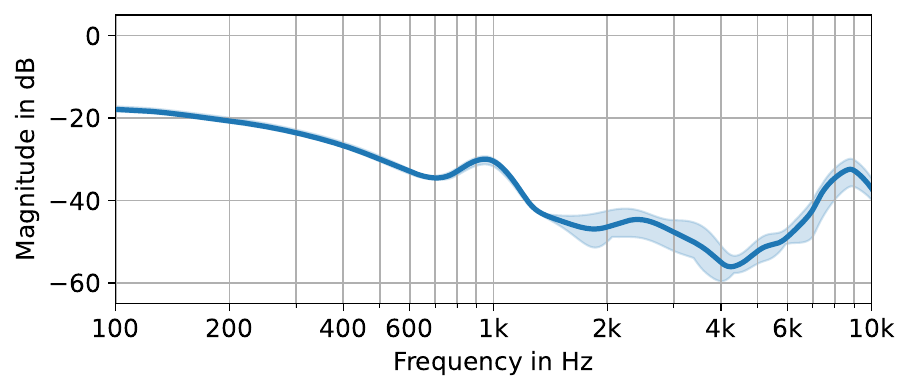}
    \caption{IL of the custom headphones. Mean in bold, std in shade.}
    \label{fig:il_meas}
\end{figure}
The results of the insertion loss measurements described in Section \ref{eval:inherent-il} are shown in Figure \ref{fig:il_meas}. Above approximately $150$Hz, the headphones exhibit an IL exceeding $-20$\,dB. Below that, the IL decreases slightly, but consistently remains at or below $-18$\,dB over the entire frequency range. This is in line with the design requirements, exceeding the defined target. Given these results and maintaining a separation of at least $10$\,dB between the emulated and the leakage components, the system is able to effectively emulate ILs down to $-20$\,dB and more, depending on the exact OE target curve. Thus, neglecting the leakage component in the filter derivation is justified.
\subsection{Inherent Occlusion Gain}
\label{subsec:res_OG}
\begin{figure}[t]
    \centering
    \includegraphics[width=0.65\linewidth]{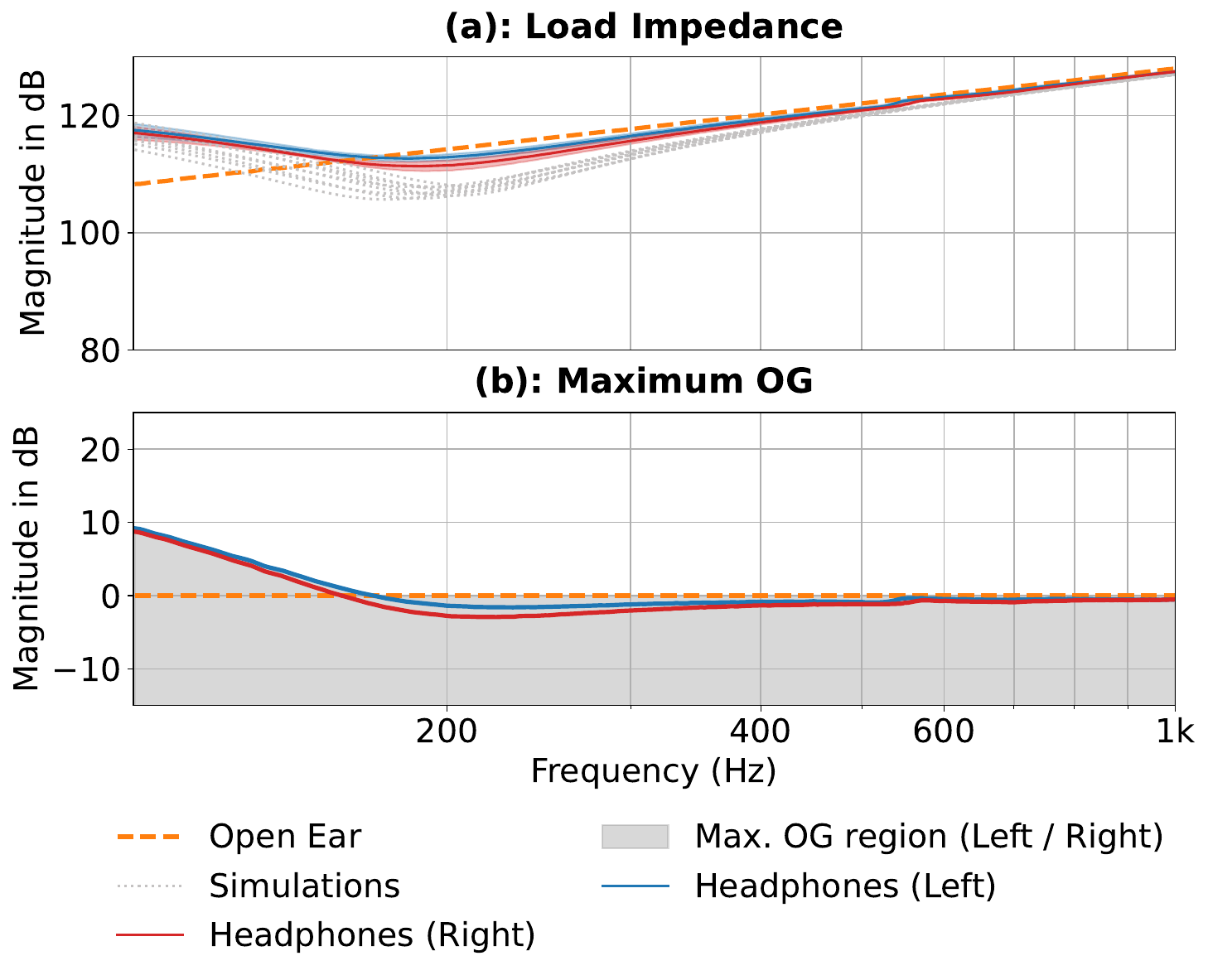}
    \caption{Measured load impedance (a) and maximum OG (b). Mean in bold, std in shade.}
    \label{fig:imp_og_meas}
\end{figure}

The results of the impedance measurements described in Section \ref{eval:inherent-og} are shown in Figure \ref{fig:imp_og_meas}. Position (a) shows the magnitude of the measured load impedance in dB with the open ear radiation impedance as reference in orange clearly exhibiting the expected mass-like behaviour. The load impedances are shown in blue for the left and red for the right side with negligible differences visible between the two. Furthermore, the blue and red shaded regions visualising the standard deviation around the means never exceed $2.26$\,dB, showing low variability between individual measurements. These results highlight the consistency between both the two sides and subsequent measurements, validating the use of 3D printing for the construction of the headphones.

The grey dotted lines show the analytical estimations of the load impedance based on the transmission line model for different geometry parameter permutations. For more information on the model and the parameters, see Appendix \ref{app:tl_model}. Measurements and simulations generally are in good agreement and exhibit very similar resonance frequencies, indicating that the reactive parameters of the system are accurately modelled. However, the simulations predict a slightly steeper resonance notch with a lower minimum impedance magnitude compared to the empirical data. This can be explained by a slight under-estimation of the system's damping in the simulations, which is likely due to the assumptions made regarding the material parameters. Nevertheless, the built system outperforms the simulations over the entire frequency range, confirming the validity of the simplified transmission line model for the design of the headphones.

When comparing the load impedances to the open ear radiation impedance, accurate matching with a maximum deviation below $3$\,dB is observed above approximately $150$\,Hz. Above $400$\,Hz the maximum deviation decreases further to below $1.4$\,dB. Below $150$\,Hz, however, the load impedance rises above the reference due to the compliance behaviour of the enclosed volume dominating in this frequency range. Nevertheless, its magnitude is still far smaller than $Z_{\text{down}}$, which \citet[Figure 6b]{carilloTheoreticalInvestigationLow2020} estimates to be approximately $165$\,dB at $100$\,Hz. Thus, omitting $Z_{\text{down}}$ in the approximation of $Z_{\text{tot}}^{\text{occl, HP}}$ in Equation \ref{eq:og_occ_approx} is justified.

\begin{figure}[b]
    \centering
    \includegraphics[width=0.65\linewidth]{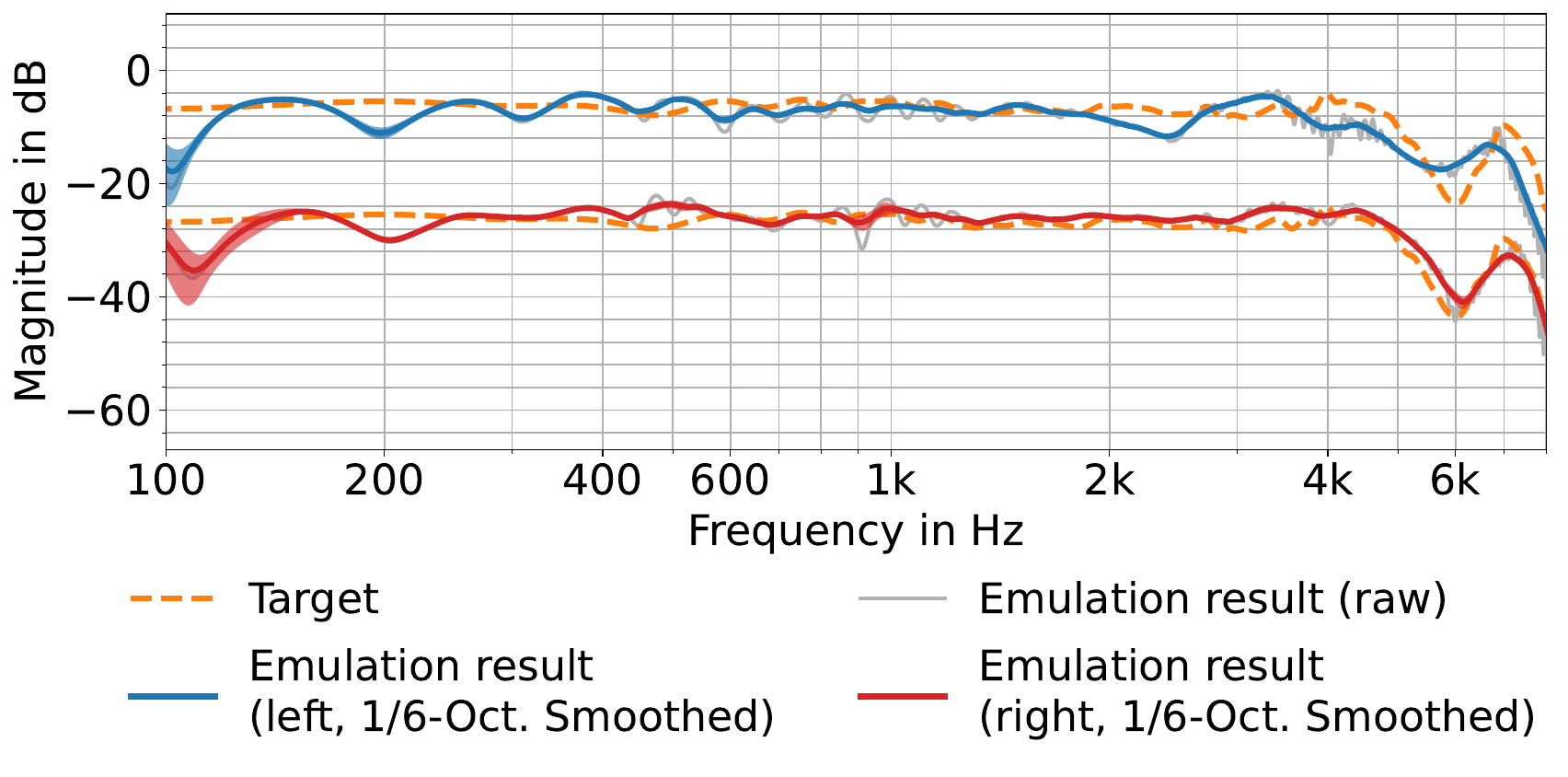}
    \caption{Measured transfer functions with active real-time filtering, shown raw (grey) and 1/6-octave smoothed (blue and red), compared to the target transfer function (orange) for the $0$\,dB OE target. Target and results for the right side shifted down by $20$\,dB for better visibility. Mean in bold, std in shade.}
    \label{fig:transp_emulation_results}
\end{figure}
Position (b) of Figure \ref{fig:imp_og_meas} shows the difference between $Z_{\text{load}}$ and $Z_{\text{open}}$ (right side in red, left in blue) in dB, which when positive serves as an upper bound for the headphones' inherent OG according to Equation \ref{eq:og_final_approx}.
The grey shaded area shows the full region in which the inherent OG is guaranteed to lie according to Equation \ref{eq:og_final_approx}.
The results show an absolute maximum OG of $8.7$\,dB at $100$\,Hz.
Above that, the upper bound for the OG decreases rapidly, reaching $0$\,dB at around $155$\,Hz.
From thereon, the OG is guaranteed to be below $0$\,dB.
As the BC component of the own voice is already assumed to be perceptually irrelevant at $0$\,dB OG compared to the AC component \cite{porschmannInfluencesBoneConduction2000}, we can deduce that the proposed design does not alter the perception of the own voice BC component above $155$\,Hz, which is close to the mean fundamental frequency of the human voice \cite{yangFundamentalFrequencyF02024}.
Despite that, a slight but perceptible OG for users with low voices might remain.
This is most applicable to male voices, who tend to have lower fundamental frequencies (mean: $116$\,Hz \cite{yangFundamentalFrequencyF02024}).
At these frequencies, the inherent OG is still guaranteed to be below $6$\,dB which is far below typical OGs that are to be emulated with this system.
Thus, this limitation was accepted as a trade-off to keep the headphones comparatively light and self-contained. Nevertheless, perceptual studies using this system should test and potentially exclude participants with voices below $110$\,Hz during the screening process, as the residual OG in that range could become problematic depending on the exact study design.
\begin{figure*}[t]
    \centering
    \includegraphics[width=\linewidth]{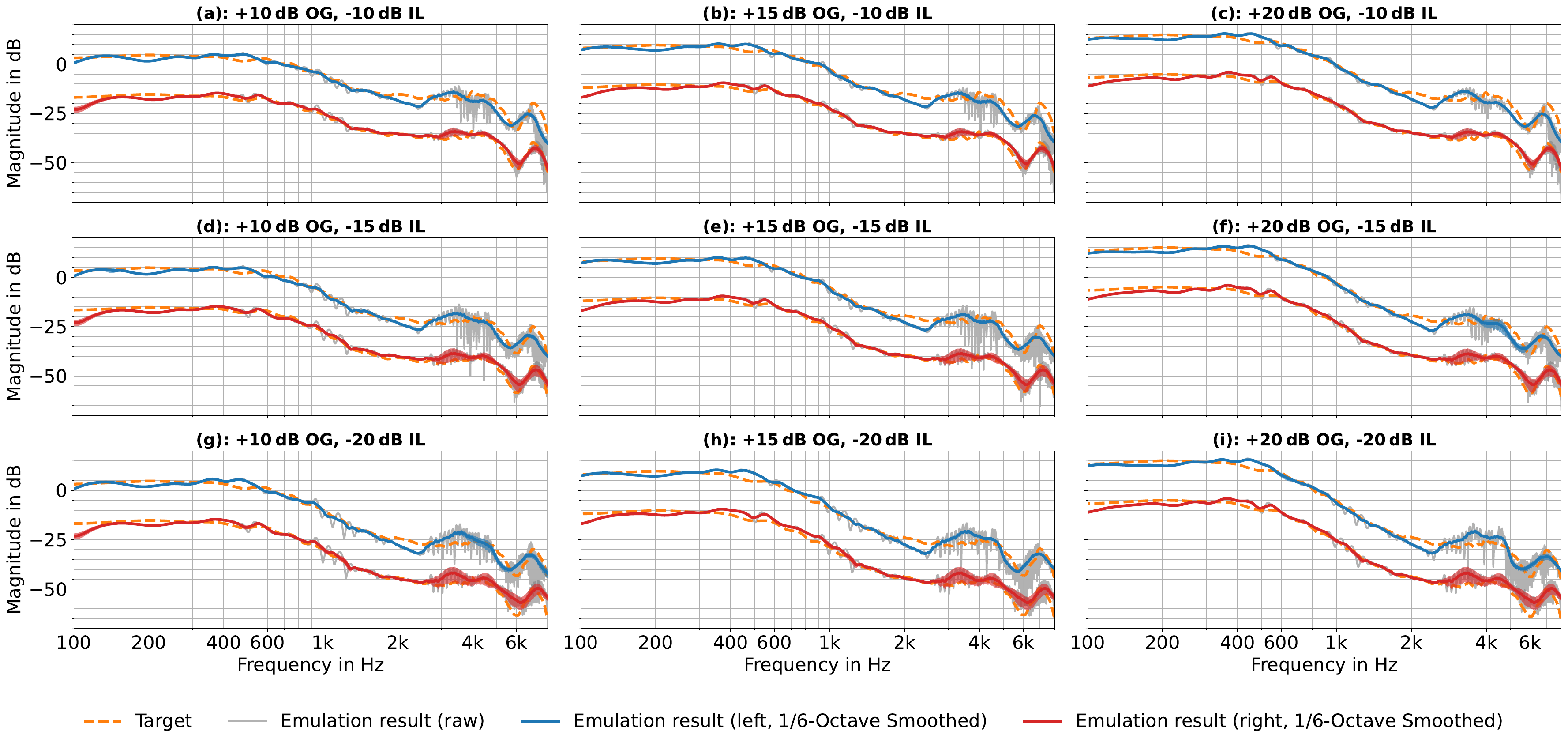}
    \caption{Measured transfer functions with active real-time filtering, shown raw (grey) and 1/6-octave smoothed (blue and red), compared to the respective target transfer function (orange) for each non-zero OE target. Target and results for the right side shifted down by $20$\,dB for better visibility. Mean in bold, std in shade.}
    \label{fig:oe_emulation_results}
\end{figure*}
\subsection{OE Emulation}
\label{subsec:res_emul}
Figure \ref{fig:transp_emulation_results} shows the measured emulation results for the $0$\,dB target, which corresponds to the open ear condition. The system's transfer function under the emulation filter is shown in blue for the left and red for the right ear, the open ear target transfer function is shown in orange. 
Overall, the results show good matching between the emulated and target transfer functions. Some comb filtering can be observed until around $700$\,Hz, stemming from the superposition of the reproduced signal and the leakage component. These ripples become progressively smaller with increasing frequency, in line with the increasing insertion loss shown in Figure \ref{fig:il_meas}. An exception to this is a more pronounced ripple peak around $1$\,kHz, which coincides with the resonance peak apparent around $1$\,kHz in Figure \ref{fig:il_meas} and can thus be attributed to a slight increase in the leakage component due to cavity resonances of the headphone system. The left side exhibits more pronounced comb filtering in the low frequency regime, suggesting a higher influence of the leakage component compared to the right side. Additionally, the right side shows a better matching to the target transfer function above $2$\,kHz. Both can be attributed to a better fit of the headphones on the right side of the artificial head, which is also apparent from the lower variance of the measurements. This is a limitation of the measurement approach, as the hard surface of the artificial head and its narrowing neck geometry limit the ability to achieve a good seal between head and earcups on both sides simultaneously. Generally, the results show that the system is able to match the open ear transfer function reasonably well, which, while not the core goal of this work, is a prerequisite for the emulation of arbitrary OE targets. With the emulation, the severity of the comb filtering effect decreases further as the additional low-frequency gain of the OG emulation increases the level distance to the leakage component, thus reducing its influence on the overall system response. This can be seen in the emulation results for the nine different OE targets.

Figure \ref{fig:oe_emulation_results} shows the results for all investigated OE targets, grouped horizontally by IL and vertically by OG. Orange again depicts the target transfer function, while the measured transfer functions of the system with real-time filtering enabled are shown in blue (left) and red (right). Compared to the results for the $0$\,dB case, the comb filtering artefacts are already significantly decreased at $+10$\,dB OG. Analysing the plots from lower to higher OGs (left to right) also shows the comb filtering to become progressively less pronounced for frequencies below roughly $1$\,kHz with increasing OG.
From the results it is apparent that the system is able to match arbitrary combinations of IL and OG targets given they are combined into appropriate OE target curves (Figure \ref{fig:oe_targets}).

The variance of the emulation does, however, increase from around $2.5$\,kHz onwards in all measurements. In Figure \ref{fig:oe_emulation_results}, this is apparent in the unsmoothed measurement results shown in grey.
This behaviour coincides with the transition to higher-order modes in the extender sections: Given a cross-sectional area of approximately $60$\,cm² for the largest part of the extenders (see Appendix \ref{app:tl_model}), the cutoff frequency for plane wave propagation according to ISO 10534-2 \cite{isoAbsorbtionImpedanceTubes2023} is around $2.3$\,kHz, above which higher-order modes are able to propagate.
Furthermore, the damping material filling the extenders influences higher frequencies more severely, especially at low reproduction levels with less sound power to overcome the heavy internal damping, adding to the observed variance.
Lastly, the variance seems to increase with increasing IL (top to bottom), which can be explained by the reduced level of the reproduced signal, thus raising the relative contribution of the leakage component and consequently the comb filtering effect.
Despite this, the system still tracks the target well even for these higher frequencies as shown by the $1/6$-octave filtered results in Figure \ref{fig:oe_emulation_results}.
As $1/6$-octave filtering approximates the frequency resolution of the human ear above $2$\,kHz reasonably
well, we conclude that the system is able to sufficiently match the target transfer functions for all OE targets for the intended application, proving the efficacy of the proposed approach to emulate and independently control IL and OG.

\subsection{Limitations and Suggestions for Future Work}
\label{subsec: limits-and-outlook}
It must be noted that the proposed approach relies on several assumptions and simplifications.
For the estimation of the headphone-inherent OG, we assumed that BC transmission to the outer ear is unaltered by the presence of the headphones, allowing the OG to be expressed solely through the involved impedances.
While insufficient to fully describe the mechanics of the OE given arbitrary occlusion devices \cite{kerstenImpactEarCanal2024}, this assumption is valid for the derivation of an upper bound for the headphones' inherent OG. This is because the headphones are positioned over the ears instead of inside the ear canal and thus expected to influence the ear canal wall vibrations even less than typical occlusion devices.

Secondly, we could not achieve a perfect match between the headphones' load impedance and the open ear radiation impedance down to $100$\,Hz while still maintaining a lightweight, self-contained form factor for the headphones. This results in a slight but perceptible inherent OG for low voices, which could be problematic for perceptual studies. Thus, we recommend screening participants and potentially excluding those with voices below $110$\,Hz depending on the exact study design to avoid any perceptual influence of the residual OG.

Furthermore, while the physical OE varies depending on the spoken phoneme \cite{blauMethodsExperimentallyCharacterize2025}, we assumed this time-variant behaviour to be of secondary perceptual importance and focused on a time-invariant model.
Despite that, the presented approach has the potential to elucidate the perceptual relevance of this time-variant nature, which is currently unclear \cite{blauMethodsExperimentallyCharacterize2025}. Future research should thus integrate time-variant effects into the emulation and evaluate their perceptual impact through formal listening tests.

Lastly, the perception of one's own voice is governed not only by the level differences between the individual AC and BC components, but also by their phase interactions. As all AC and BC components have different propagation delays and ultimately superpose in the cochlea, they create comb filtering artefacts yielding the specific timbre of one's own voice. Hence we expect a slight alteration in own voice timbre despite the accurate emulation of the OE conditions in level due to the involved system delay. Based on informal subjective evaluations conducted so far, we expect this to be of secondary importance, especially since the OE is primarily a level-based phenomenon and BC propagation to the middle and inner ear, which is important to own voice perception \cite{porschmannInfluencesBoneConduction2000}, remains unaltered by the headphones. In future work, the OE emulation DSP chain should nevertheless be implemented on dedicated low-latency hardware to further reduce this unintended change in own voice perception.

Overall, the results show the efficacy of the proposed approach to emulate and independently control repeatable IL and OG conditions, laying the groundwork for in-depth investigations into the perceptual principles underlying the OE through formal listening tests. A perceptual evaluation through application of the approach in such a study is thus the next logical step in this research.

\section{Conclusion}
\label{sec:conclusion}
This work aimed to develop an approach for precisely controlling the occlusion effect (OE) experienced by participants in perceptual studies while also allowing for the independent manipulation of its insertion loss (IL) and occlusion gain (OG) components. To achieve this, a custom headphone system was designed based on commercially available earmuffs to emulate the OE by reproducing an appropriately filtered version of the user's speech in real-time.
We evaluated the system's capability to emulate OGs of $10$\,dB, $15$\,dB, and $20$\,dB, ILs of $-10$\,dB, $-15$\,dB, and $-20$\,dB, as well as a $0$\,dB target representing the open-ear condition.
Overall, the system sufficiently matched the target transfer functions, demonstrating the efficacy of this approach for the independent emulation and control of IL and OG. Limitations arise mainly from a slight but perceptible inherent OG of the headphones for deep voices and the processing delay potentially altering the perceived timbre of the own voice. Overall, this work lays the foundation for future perceptual studies investigating the influence of the OE on own voice perception and communication. Future research will focus on a perceptual evaluation of the system through application in a formal listening test, its extension to time-variant OE emulation, and the implementation of the DSP chain on dedicated low-latency hardware.

\acknowtext

Large language models have been used to assist in the writing and editing process of this manuscript. After usage, the authors reviewed and edited the content as needed and take full responsibility for the content of the published article.

The authors would like to thank Xiao Lou for his contributions to the supplemental material of this publication. In particular, he consolidated the preprocessing and analysis scripts into a unified repository and improved their readability and structure for publication.

\funding

This work was partly supported by the Deutsche Forschungsgemeinschaft (DFG, German Research Foundation) - Project-ID 352015383 - SFB 1330.

\conflict

The authors have no conflict of interest to disclose.

\dataavailability

The 3D-printable geometry files, data, and analysis scripts supporting the findings of this work are currently available at \url{https://rwth-aachen.sciebo.de/s/QfdHF7FyJJdcxoG}. Upon final acceptance, this link will be replaced by a Zenodo deposit.

\authorcontrib

\textbf{R. Rehman:} Conceptualization, Data curation, Formal analysis, Investigation, Methodology, Software, Visualization, Writing -- original draft.
\textbf{S. Kersten:} Conceptualization, Investigation, Methodology, Writing -- review \& editing.
\textbf{A. Schliep:} Investigation, Methodology, Visualization, Writing -- review \& editing.
\textbf{J. Fels:} Conceptualization, Project administration, Funding acquisition, Supervision, Writing -- review \& editing.




\bibliography{bibliography}
\bibliographystyle{acta}


\appendix

\section{Transmission Line Model}
\label{app:tl_model}
\begin{figure}
    \centering
    \includegraphics[width=0.65\linewidth]{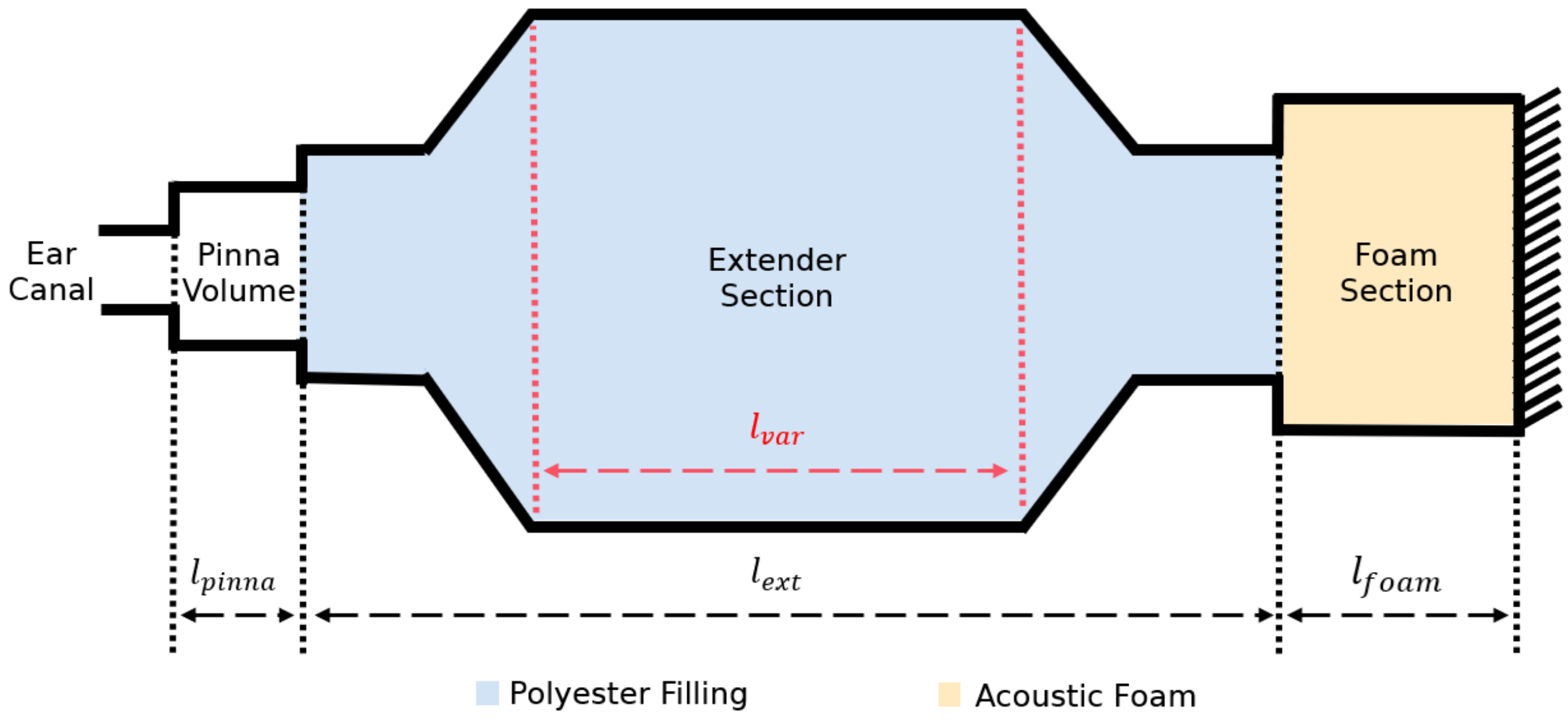}
    \caption{Transmission line model of the headphones' inner extender section.}
    \label{fig:tl_model}
\end{figure}

\begin{figure}
    \centering
    \includegraphics[width=0.65\linewidth]{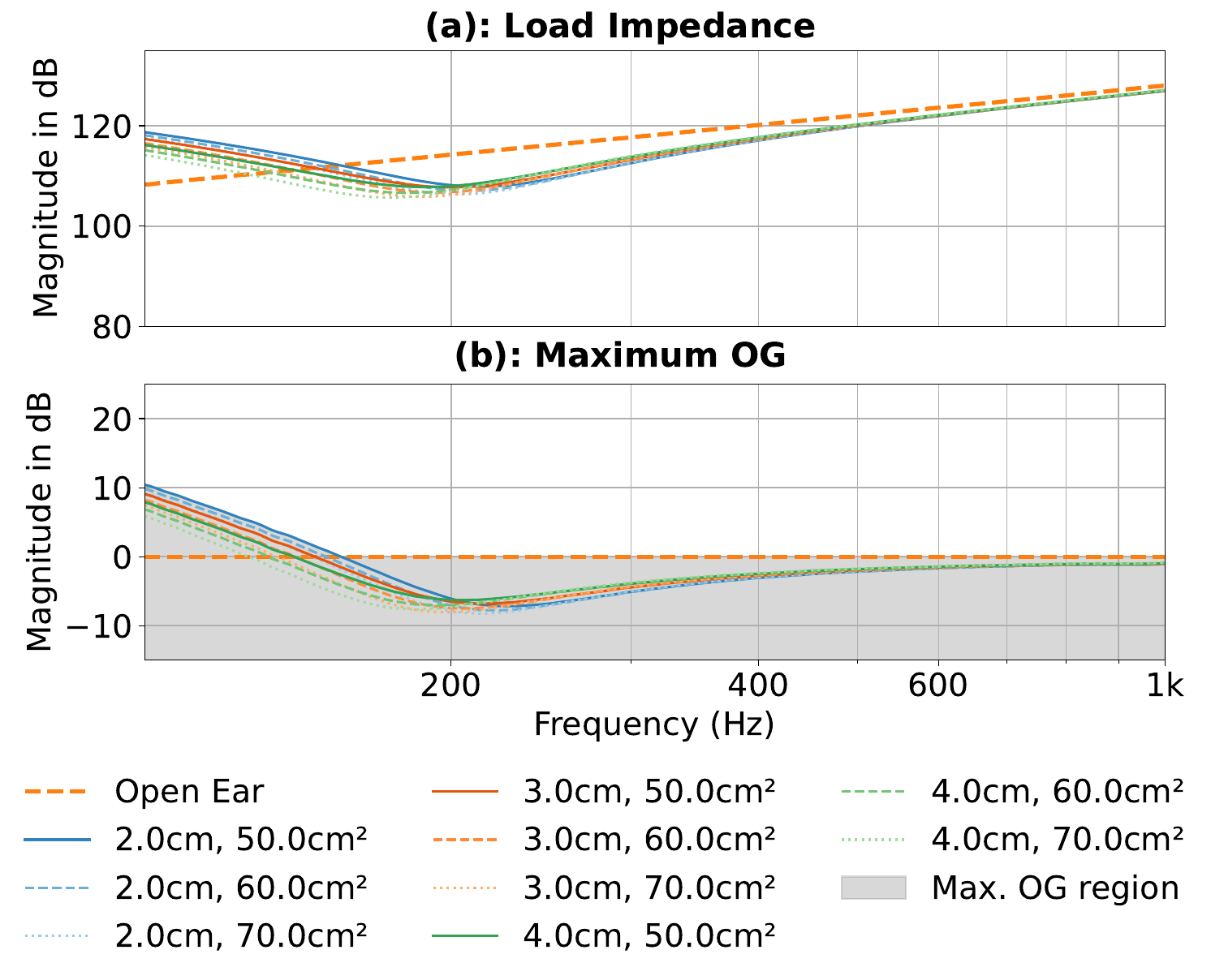}
    \caption{Simulated load impedance of the headphone extender section for different geometry parameter permutations (top) and the corresponding maximum OG inherent to the headphones (bottom).}
    \label{fig:impedance_sim}
\end{figure}

Figure \ref{fig:tl_model} shows the transmission line model of the inner headphone extender section. The model consists of a series of connected transmission lines and conical horns filled with polyester filling. The earmuff capsule was modelled as a transmission line filled with acoustic foam and terminated with a rigid termination. To facilitate a good connection of the extender to the earcup at one end and to the earmuff capsule at the other, we kept the geometries of the extender ends constant and varied the cross-sectional area and length of the middle section marked in red. All simulations were conducted in Python using the Pyfar library \cite{brinkmannOpenEducationalResources2025}. The polyester filling and acoustic foam were modelled using the Johnson-Champoux-Allard equivalent fluid model \cite{allardPropagationSoundPorous2009}. The parameters for air at room temperature were taken from \citet{carilloReductionOcclusionEffect2022}. For the material parameters of the acoustic foam in the earmuff capsule, we used those reported by \citet{carilloReductionOcclusionEffect2022} as a starting point and adjusted the air flow resistivity until the damping characteristics matched initial impedance measurements. All parameters are summarised in Table \ref{tab:jca-air-params} and \ref{tab:jca-material-params}. The resulting load impedances and maximum OG according to Equation \ref{eq:og_final_approx} in dependence on the geometry configuration are shown in Figure \ref{fig:impedance_sim}. Based on these results, we settled for the ($3$\,cm, $60$\,cm²)-configuration, as this yielded the best compromise between inherent OG and physical size and weight of the headphones when taking into account the full geometry. Due to practical limitations during the CAD design process given the earmuffs' oval shape, the final dimensions deviated slightly from this target. The final inner section of the inner extender has a length of $3$\,cm and a cross-sectional area of $59.1$\,cm². In total, the inner extender section has a length of $11.5$\,cm, with its cross-sectional area varying between $30$\,cm² and $59.1$\,cm².
\begin{table}[h]
    \centering
    \begin{tabular}{l|l|l|l|l|l}
        $\rho_0$ [kg/m$^3$] & $c_0$ [m/s] & $\eta$ [Pa s]      & $\gamma$ [-] & $C_p$ [J/kg/K]   & $\kappa$ [W/m/K] \\
        \toprule
        $1.2$               & $343$       & $1.83\cdot10^{-5}$ & $1.4$        & $1.002\cdot10^3$ & $0.025$
    \end{tabular}
    \caption{Properties of the air at room conditions.}
    \label{tab:jca-air-params}
    \begin{tabular}{l|l|l|l|l|l}
        Material          & $\Phi$ [-] & $\alpha_\infty$ [-] & $\sigma$ [Ns/m$^4$] & $\Lambda$ [m]     & $\Lambda'$ [m]     \\
        \toprule
        Polyester Filling & $0.95$     & $1.0$               & $10,000$            & $1\cdot10^{-4}$   & $2\cdot10^{-4}$    \\
        Acoustic Foam     & $0.9$      & $1.1$               & $15,000$            & $8.7\cdot10^{-5}$ & $1.63\cdot10^{-4}$
    \end{tabular}
    \caption{Properties of the polyester filling and acoustic foam.}
    \label{tab:jca-material-params}
\end{table}
\clearpage
\section{Least-Squares Optimisation}
\label{app:regul-solution}
Let $\mathbf{h}_{\text{ac}}^{\text{open}}$ be the impulse response (IR) vector of length $L_o$ corresponding to $H_{\text{ac}}^{\text{open}}$ and $\mathbf{h_{oe}}$ the IR vector of length $L_\text{oe}$ corresponding to $OE$. Let $\mathbf{g}$ be the IR vector of length $L_g$ corresponding to $G$, $\mathbf{h}_{\text{ext}}$ the IR vector of length $L_\text{ext}$ corresponding to $H_{\text{ext}}$, and $\mathbf{h}_{\text{mic}}$ the IR vector of length $L_\text{mic}$ corresponding to $H_{\text{mic}}$.

Given two IR vectors $\mathbf{h}$ and $\mathbf{x}$ of length $L_h$ and $L_x$ respectively, a convolution matrix $\mathbf{H}$ is a $(L_x+L_h-1)\times L_x$ Toeplitz matrix of the corresponding IR vector $\mathbf{h}$, such that the matrix-vector-product $\mathbf{H}\mathbf{x}$ equals the convolution of $\mathbf{h}$ and $\mathbf{x}$.
With the convolution matrices $\mathbf{H}_{\text{ext}}$ and $\mathbf{H}_{\text{mic}}$ for $\mathbf{h}_{\text{ext}}$ and $\mathbf{h}_{\text{mic}}$, and $\mathbf{OE}$ being the zero-padded convolution matrix of $\mathbf{h}_{oe}$, the problem from Equation \ref{eq:main_G} can be expressed as:
\begin{equation}
    \label{eq:appendix-1}
    \mathbf{H}_{\text{mic}}\mathbf{H}_{\text{ext}}\mathbf{g} = \mathbf{OE} \text{ } \mathbf{h}_{\text{ac}}^{\text{open}}
\end{equation}
Zero-padding of $\mathbf{OE}$ is necessary to match the dimensions of $\mathbf{H}_{\text{mic}}\mathbf{H}_{\text{ext}}$, as $L_\text{oe} + L_\text{o} - 1 < L_\text{mic}+L_\text{ext}+L_g-2$ in practice. The involved matrices are
    {\allowdisplaybreaks
        \begin{align*}
            \mathbf{H}_{\text{ext}} & = \overset{\xrightarrow{\hspace{5em}L_g\hspace{5em}}\hspace{5em}}{\left.
            \begin{bmatrix}
                    h_{\text{ext},0}              & 0                             & \cdots & 0                             \\
                    h_{\text{ext},1}              & h_{\text{ext},0}              & \ddots & \vdots                        \\
                    \vdots                        & \ddots                        & \ddots & 0                             \\
                    \vdots                        & \ddots                        & \ddots & h_{\text{ext},0}              \\
                    \vdots                        & \ddots                        & \ddots & h_{\text{ext},1}              \\
                    h_{\text{ext},L_\text{ext}-1} & h_{\text{ext},L_\text{ext}-2} & \ddots & \vdots                        \\
                    0                             & h_{\text{ext},L_\text{ext}-1} & \ddots & \vdots                        \\
                    \vdots                        & \ddots                        & \ddots & \vdots                        \\
                    0                             & 0                             & \cdots & h_{\text{ext},L_\text{ext}-1}
                \end{bmatrix} \right\downarrow_{L_\text{ext}+L_g-1}}                                                                                                                  \\[3ex]
            \mathbf{H}_{\text{mic}} & = \overset{\xrightarrow{\hspace{4em}L_\text{ext}+L_g-1\hspace{4em}}\hspace{7em}}{\left.\begin{bmatrix}
                                                                                                                                     h_{\text{mic},0}              & 0                             & \cdots & 0                             \\
                                                                                                                                     h_{\text{mic},1}              & h_{\text{mic},0}              & \ddots & \vdots                        \\
                                                                                                                                     \vdots                        & \ddots                        & \ddots & 0                             \\
                                                                                                                                     \vdots                        & \ddots                        & \ddots & h_{\text{mic},0}              \\
                                                                                                                                     \vdots                        & \ddots                        & \ddots & h_{\text{mic},1}              \\
                                                                                                                                     h_{\text{mic},L_\text{mic}-1} & h_{\text{mic},L_\text{mic}-2} & \ddots & \vdots                        \\
                                                                                                                                     0                             & h_{\text{mic},L_\text{mic}-1} & \ddots & \vdots                        \\
                                                                                                                                     \vdots                        & \ddots                        & \ddots & \vdots                        \\
                                                                                                                                     0                             & 0                             & \cdots & h_{\text{mic},L_\text{mic}-1}
                                                                                                                                 \end{bmatrix} \right\downarrow_{L_\text{mic}+L_\text{ext}+L_g-2}} \\[3ex]
            \mathbf{OE}             & = \overset{\xrightarrow{\hspace{5em}L_o\hspace{5em}}\hspace{5em}}{\left.
                \begin{bmatrix}
                    h_{oe, 0}             & 0                     & \cdots & 0                     \\
                    h_{oe, 1}             & h_{oe, 0}             & \ddots & \vdots                \\
                    \vdots                & \ddots                & \ddots & 0                     \\
                    \vdots                & \ddots                & \ddots & h_{oe, 0}             \\
                    \vdots                & \ddots                & \ddots & h_{oe, 1}             \\
                    h_{oe, L_\text{oe}-1} & h_{oe, L_\text{oe}-2} & \ddots & 0                     \\
                    0                     & h_{oe, L_\text{oe}-1} & \ddots & 0                     \\
                    \vdots                & \ddots                & \ddots & \vdots                \\
                    0                     & 0                     & \ddots & h_{oe, L_\text{oe}-1} \\
                    0                     & 0                     & \ddots & 0                     \\
                    \vdots                & \ddots                & \ddots & \vdots                \\
                    0                     & 0                     & \cdots & 0
                \end{bmatrix} \right\downarrow_{L_\text{mic}+L_\text{ext}+L_g-2}}
        \end{align*}}

As there is no exact solution for Equation \ref{eq:appendix-1}, we can instead find the optimal solution $\mathbf{g}^{opt}$ by minimising the cost function $J(\mathbf{g})$:
\begin{align}
    J(\mathbf{g}) 
     & \coloneq \lVert \mathbf{A}\mathbf{g} - \mathbf{v} \rVert^2_2
\end{align}
with $\mathbf{A} = \mathbf{H}_{\text{mic}}\mathbf{H}_{\text{ext}}$ and $\mathbf{v} = \mathbf{OE}\text{ }\mathbf{h}_{\text{ac}}^{\text{open}}$.
The optimal solution $\mathbf{g}^{opt}$ is the vector whose matrix-vector product with $\mathbf{A}$ yields the orthogonal projection of $\mathbf{v}$ onto the column space of $\mathbf{A}$:
\begin{align*}
    \operatorname{proj}_{C(\mathbf{A})}(\mathbf{v}) & = \mathbf{A}\mathbf{g}^{opt}                          \\
    \mathbf{g}^{opt}                                & = (\mathbf{A}^T\mathbf{A})^{-1}\mathbf{A}^T\mathbf{v}
\end{align*}
This matrix inversion is, however, typically ill-conditioned. To control for this, we include a regularisation term \cite{schepkerRobustSingleMultiloudspeaker2022}:
\begin{align*}
    J_{\text{regul}}(\mathbf{g}) & = \lVert \mathbf{A}\mathbf{g} - \mathbf{v} \rVert^2_2 + \lambda \lVert \mathbf{g} \rVert^2_2
\end{align*}
With $\mathbf{I}$ being the identity matrix, the optimal solution becomes:
\begin{align*}
    \mathbf{g}^{opt} & = (\mathbf{A}^T\mathbf{A} + \lambda \mathbf{I})^{-1}\mathbf{A}^T\mathbf{v}                                                                                                   \\
                     & = \left[\left( \mathbf{H_{\text{mic}}} \mathbf{H_{\text{ext}}} \right)^\top \left( \mathbf{H_{\text{mic}}} \mathbf{H_{\text{ext}}} \right) + \lambda \mathbf{I} \right]^{-1} \\
                     & \quad \left( \mathbf{H_{\text{mic}}} \mathbf{H_{\text{ext}}} \right)^\top \left( \mathbf{OE}\text{ }\mathbf{h_{\text{ac}}^{\text{open}}} \right)
\end{align*}

\end{document}